\documentclass[preprint,12pt]{article}
\pdfoutput=1

\usepackage{amsmath,amssymb,array,calc,rotating,epsfig,psfrag, amscd, datetime}

\usepackage{color}
\usepackage[
      colorlinks=true,
      urlcolor=blue,    
      filecolor=blue,     
      citecolor=red,
      pdfstartview=FitV,
       bookmarksopen=true    
      ]{hyperref}
\usepackage[left=2cm,top=2cm,right=2cm,nohead]{geometry}
\numberwithin{equation}{section}

\newcommand{\nc}{\newcommand}

\definecolor{cardinal}{rgb}{0.6,0,0}
\definecolor{darkgreen}{rgb}{0,0.5,0}
\definecolor{golden}{rgb}{0.92, 0.7, 0}
\definecolor{midnight}{rgb}{0, 0, 0.5}
\definecolor{darkblue}{rgb}{0.2, 0, 0.8}

\nc{\ra}{\rightarrow} 
\nc{\lra}{\leftrightarrow} 
\nc{\Ra}{\Rightarrow} 
\nc{\LRa}{\Leftightarrow} 
\nc{\blp}{{\big (}}
\nc{\brp}{{\big )}}
\nc{\Blp}{{\Big (}}
\nc{\Brp}{{\Big )}}
\nc{\bglp}{{\bigg (}}
\nc{\bgrp}{{\bigg )}}
\nc{\Bglp}{{\Bigg (}}
\nc{\Bgrp}{{\Bigg )}}
\nc{\slb}{{\rm [}}
\nc{\srb}{{\rm ]}}
\nc{\bslb}{{\rm \big [}}
\nc{\bsrb}{{\rm \big ]}}
\nc{\Bslb}{{\rm \Big [}}
\nc{\Bsrb}{{\rm \Big ]}}

\def\al{\alpha}

\def\eps{\epsilon}
\nc{\veps}{\varepsilon}
\def\gam{\gamma}

\def\lam{\lambda}
\def\om{\omega}

\nc{\vphi}{\varphi}
\def\tha{\theta}

\def\sig{\sigma}

\def\Gam{\Gamma}

\def\Lam{\Lambda}
\def\Om{\Omega}
\def\Sig{\Sigma}

\def\coeff#1#2{\relax{\textstyle {#1 \over #2}}\displaystyle}

\nc{\myvspace}{\rule[-1em]{0pt}{2.5em}}
\nc{\bea}{\begin{eqnarray}}
\nc{\eea}{\end{eqnarray}}
\nc{\be}{\begin{equation}}
\nc{\ee}{\end{equation}}
\nc{\barr}{\begin{array}}
\nc{\earr}{\end{array}}
\nc{\co}{{\cal o}}

\nc{\cA}{{\cal A}}
\nc{\cB}{{ \cal B}}
\def\cC{{\cal C}}
\def\cD{{\cal D}}

\nc{\cF}{{\cal F}}
\nc{\cG}{{\cal G}}

\def\cI{{\cal I}}
\def\cJ{{\cal J}}
\def\cK{{\cal K}}
\nc{\cL}{{\cal L}}
\nc{\cM}{{\cal M}}

\def\cN{{\cal N}}

\def\cO{{\cal O}}
\def\cP{{\cal P}}
\nc{\cQ}{{\cal Q}}
\nc{\cR}{{\cal R}}
\def\cS{{\cal S}}
\def\cT{{\cal T}}

\def\cV{{\cal V}}
\def\cV{{\cal V}}
\def\cW{{\cal W}}

\def\cZ{{\cal Z}}
\nc{\cQd}{\cQ^{\dagger}}
\nc{\cRd}{\cR^{\dagger}}
\nc{\BB}{{\mathbb B}}
\nc{\CC}{{\mathbb C}}
\nc{\DD}{{\mathbb D}}
\nc{\EE}{{\mathbb E}}
\nc{\FF}{{\mathbb F}}
\nc{\GG}{{\mathbb G}}
\nc{\HH}{{\mathbb H}}
\nc{\JJ}{{\mathbb J}}
\nc{\MM}{{\mathbb M}}
\nc{\RR}{{\mathbb R}}
\nc{\PP}{{\mathbb P}}
\nc{\QQ}{{\mathbb Q}}
\nc{\UU}{{\mathbb U}}
\nc{\ZZ}{{\mathbb Z}}
\nc{\calone}{{\mathbb 1}}
\nc{\half}{\coeff{1}{2}}
\nc{\quarter}{\coeff{1}{4}}
\nc{\del}{\partial}

\nc{\delbar}{\bar\partial}
\nc{\thalf}{\frac{t}{2}}
\nc{\Spin}{\operatorname{Spin}}
\nc{\SO}{\operatorname{SO}}

\nc{\Sp}{{\rm Sp}}
\nc{\com}[2]{{ \left[ #1, #2 \right] }}
\nc{\acom}[2]{{ \left\{ #1, #2 \right\} }}
\nc{\rr}{\rightarrow}
\nc{\p}{\partial}
\nc{\LT}{{\LL_\T}}
\nc{\Tr}{{\rm Tr}}
\nc{\tr}{{\rm tr}}
\nc{\Adag}{A^{\dagger}}
\nc{\AdagI}{A^{\dagger I}}
\nc{\AdagJ}{A^{\dagger J}}
\nc{\AdagK}{A^{\dagger K}}
\nc{\AdagL}{A^{\dagger L}}
\nc{\AdagM}{A^{\dagger M}}
\nc{\Bdag}{B^{\dagger}}
\nc{\BdagI}{B^{\dagger}_I}
\nc{\BdagJ}{B^{\dagger}_J}
\nc{\BdagK}{B^{\dagger}_K}
\nc{\BdagL}{B^{\dagger}_L}
\nc{\BdagM}{B^{\dagger}_M}
\nc{\Cdag}{C^{\dagger}}
\nc{\CdagI}{C^{\dagger I}}
\nc{\CdagJ}{C^{\dagger J}}
\nc{\CdagK}{C^{\dagger K}}
\nc{\Ddag}{D^{\dagger}}
\nc{\DdagI}{D^{\dagger I}}
\nc{\DdagJ}{D^{\dagger J}}
\nc{\DdagK}{D^{\dagger K}}
\nc{\bva}{\breve{a}}
\nc{\bvb}{\breve{b}}
\nc{\bvc}{\breve{c}}
\nc{\bvd}{\breve{d}}
\nc{\bve}{\breve{e}}
\nc{\bvf}{\breve{f}}
\nc{\bvg}{\breve{g}}
\nc{\bvh}{\breve{h}}
\nc{\bvi}{\breve{i}}
\nc{\bvj}{\breve{j}}
\nc{\bvk}{\breve{k}}
\nc{\bvl}{\breve{l}}
\nc{\bvm}{\breve{m}}
\nc{\bvn}{\breve{n}}
\nc{\bvo}{\breve{o}}
\nc{\bvp}{\breve{p}}
\nc{\brvq}{\breve{q}}
\nc{\bvr}{\breve{r}}
\nc{\bvs}{\breve{s}}
\nc{\bvt}{\breve{t}}
\nc{\bvu}{\breve{u}}
\nc{\bvv}{\breve{v}}
\nc{\bvw}{\breve{w}}
\nc{\bvx}{\breve{x}}
\nc{\bvy}{\breve{y}}
\nc{\bvz}{\breve{z}}

\nc{\bvA}{\breve{A}}
\nc{\bvB}{\breve{B}}
\nc{\bvC}{\breve{C}}
\nc{\bvD}{\breve{D}}
\nc{\bvE}{\breve{E}}
\nc{\bvF}{\breve{F}}
\nc{\bvG}{\breve{G}}
\nc{\bvH}{\breve{H}}
\nc{\bvI}{\breve{I}}
\nc{\bvJ}{\breve{J}}
\nc{\bvK}{\breve{K}}
\nc{\bvL}{\breve{L}}
\nc{\bvM}{\breve{M}}
\nc{\bvN}{\breve{N}}
\nc{\bvO}{\breve{O}}
\nc{\bvP}{\breve{P}}
\nc{\bvQ}{\breve{Q}}
\nc{\bvR}{\breve{R}}
\nc{\bvS}{\breve{S}}
\nc{\bvT}{\breve{T}}
\nc{\bvU}{\breve{U}}
\nc{\bvV}{\breve{V}}
\nc{\bvcV}{\breve{\cV}}
\nc{\bvW}{\breve{W}}
\nc{\bvX}{\breve{X}}
\nc{\bvY}{\breve{Y}}
\nc{\bvZ}{\breve{Z}}

\nc{\ul}[1]{{\underline{#1}}}
\nc{\tal}{\widetilde{\alpha}}
\nc{\tbeta}{\widetilde{\beta}}
\nc{\ttha}{\tilde{\theta}}
\nc{\ttau}{\tilde{\tau}}
\nc{\tTha}{\tilde{\Theta}}
\nc{\tphi}{\tilde{\phi}}
\nc{\tsig}{\tilde{\sig}}
\nc{\tom}{\widetilde{\om}}
\nc{\tOm}{\widetilde{\Om}}
\nc{\tlam}{\widetilde{\lam}}
\nc{\tLam}{\tilde{\Lam}}
\nc{\tSig}{\widetilde{\Sig}}
\nc{\tPhi}{\tilde{\Phi}}
\nc{\tPhibar}{\ol{\tPhi}}
\nc{\tPi}{\widetilde{\Pi}}
\nc{\tpsi}{\widetilde{\psi}}
\nc{\tPsi}{\tilde{\Psi}}
\nc{\tgam}{\widetilde{\gam}}
\nc{\tGam}{\widetilde{\Gam}}
\nc{\tzeta}{\tilde{\zeta}}
\nc{\tZeta}{\tilde{\Zeta}}
\nc{\teta}{\widetilde{\eta}}
\nc{\teps}{\tilde{\eps}}
\nc{\tveps}{\tilde{\veps}}
\nc{\tEta}{\tilde{\Eta}}
\nc{\tchi}{\tilde{\chi}}
\nc{\tChi}{\tilde{\Chi}}
\nc{\txi}{\tilde{\xi}}
\nc{\tXi}{\widetilde{\Xi}}
\nc{\tnu}{\tilde{\nu}}
\nc{\tmu}{\tilde{\mu}}

\nc{\ta}{\tilde a}
\nc{\tb}{\tilde b}
\nc{\tc}{\tilde c}
\nc{\te}{\tilde e}
\nc{\tf}{\widetilde f}
\nc{\tg}{\widetilde g}
\nc{\ti}{\tilde i}
\nc{\tj}{\tilde j}
\nc{\tk}{\widetilde k}
\nc{\tl}{\tilde l}
\nc{\tm}{\widetilde m}
\nc{\tn}{\tilde n}
\nc{\tp}{\tilde{p}}
\nc{\tq}{\widetilde{q}}
\nc{\trr}{{\tilde r}}
\nc{\ts}{{\tilde s}}
\nc{\tu}{{\tilde u}}
\nc{\tv}{{\tilde v}}
\nc{\tw}{{\tilde w}}
\nc{\tx}{{\tilde x}}
\nc{\ty}{{\tilde y}}
\nc{\tz}{\tilde z}
\nc{\tA}{{\widetilde A}}
\nc{\tAbar}{{\ol \tA}}
\nc{\tB}{{\widetilde B}}
\nc{\tC}{{\widetilde C}}
\nc{\tD}{{\widetilde D}}
\nc{\tE}{{\widetilde E}}
\nc{\tF}{{\widetilde F}}
\nc{\tG}{{\widetilde G}}
\nc{\tcG}{{\widetilde \cG}}
\nc{\tH}{{\widetilde H}}
\nc{\tI}{{\widetilde I}}
\nc{\tcI}{{\widetilde \cI}}
\nc{\tJ}{{\widetilde J}}
\nc{\tJbar}{{\ol {\tilde J}}}
\nc{\tK}{{\widetilde K}}
\nc{\tL}{{\widetilde L}}
\nc{\tcL}{{\widetilde \cL}}
\nc{\tcLbar}{{\ol \tcL}}
\nc{\tM}{{\widetilde M}}
\nc{\tN}{{\widetilde N}}
\nc{\tcN}{{\widetilde \cN}}
\nc{\tP}{{\widetilde P}}
\nc{\tQ}{{\widetilde Q}}
\nc{\tR}{{\widetilde R}}
\nc{\tS}{\widetilde{S}}
\nc{\tT}{\widetilde{T}}
\nc{\tU}{\widetilde{U}}
\nc{\tUU}{\widetilde{\UU}}
\nc{\tV}{\widetilde{V}}
\nc{\tcV}{\widetilde{\cV}}
\nc{\tW}{\widetilde{W}}
\nc{\tcF}{\widetilde{{\cal F}}}
\nc{\tX}{\widetilde{X}}
\nc{\tY}{\widetilde{Y}}
\nc{\tcZ}{\tilde{\cZ}}
\nc{\tcZbar}{\ol{\tcZ}}

\nc{\ha}{\hat a}
\nc{\hb}{\hat b}
\nc{\hc}{\widehat c}
\nc{\hd}{\widehat d}
\nc{\he}{\widehat e}
\nc{\hf}{\widehat f}
\nc{\hg}{\widehat g}
\nc{\hh}{\widehat h}
\nc{\hm}{\widehat m}
\nc{\hn}{\widehat n}
\nc{\hp}{\widehat p}
\nc{\hq}{\widehat q}
\nc{\hr}{\widehat r}
\nc{\hs}{\widehat s}
\nc{\hv}{\widehat v}
\nc{\hw}{\widehat w}
\nc{\hx}{\widehat x}
\nc{\hy}{\widehat y}
\nc{\hz}{\widehat z}
\nc{\zhat}{\hat z}
\nc{\hA}{\widehat{A}}
\nc{\hB}{\widehat{B}}
\nc{\hC}{\widehat{C}}
\nc{\hD}{\widehat{D}}
\nc{\hE}{\widehat{E}}
\nc{\hF}{\widehat{F}}
\nc{\hcF}{\widehat{\cF}}
\nc{\hG}{\widehat{G}}
\nc{\hcG}{\widehat{\cG}}
\nc{\hH}{\widehat{H}}
\nc{\hI}{\widehat{I}}
\nc{\hcI}{\widehat{\cI}}
\nc{\hJ}{\widehat{J}}
\nc{\hK}{\widehat{K}}
\nc{\hL}{\widehat{L}}
\nc{\hcL}{\widehat{\cL}}
\nc{\hM}{\widehat M}
\nc{\hcM}{\widehat{\cM}}
\nc{\hN}{\widehat{N}}
\nc{\hO}{\widehat{O}}
\nc{\hcO}{\widehat{\cO}}
\nc{\hP}{\widehat{P}}
\nc{\hQ}{\widehat{Q}}
\nc{\hcQ}{\widehat{\cQ}}
\nc{\hcR}{\widehat{\cR}}
\nc{\hR}{\widehat{R}}
\nc{\hS}{\widehat{S}}
\nc{\hcS}{\widehat{\cS}}
\nc{\hT}{\widehat{T}}
\nc{\hU}{\widehat{U}}
\nc{\hV}{\widehat V}
\nc{\hcV}{\widehat \cV}
\nc{\hX}{\widehat X}
\nc{\hcZ}{\widehat \cZ}
\nc{\hcZbar}{\ol{\widehat \cZ}}

\nc{\heta}{\widehat{\eta}}
\nc{\hal}{\widehat \alpha}
\nc{\hbeta}{\widehat \beta}
\nc{\hphi}{\widehat{\phi}}
\nc{\hkap}{\hat{\kappa}}
\nc{\hchi}{\widehat{\chi}}
\nc{\hpsi}{\widehat{\psi}}
\nc{\hgam}{\widehat{\gam}}
\nc{\hPhi}{\hat{\Phi}}
\nc{\hPsi}{\hat{\Psi}}
\nc{\hGam}{\hat{\Gam}}
\nc{\omhat}{\widehat{\om}}
\nc{\htha}{\hat{\tha}}
\nc{\hrho}{\widehat{\rho}}
\nc{\hdel}{\widehat{\del}}

\nc{\w}{\wedge}

\nc{\vb}{\vec b}
\nc{\vc}{\vec c}
\nc{\vd}{\vec d}
\nc{\ve}{\vec e}
\nc{\vf}{\vec f}
\nc{\vg}{\vec g}
\nc{\vh}{\vec h}
\nc{\vp}{\vec p}
\nc{\vq}{\vec q}
\nc{\vr}{\vec r}
\nc{\vs}{\vec s}
\nc{\vv}{\vec v}
\nc{\vw}{\vec w}
\nc{\vx}{\vec x}
\nc{\vy}{\vec y}
\nc{\vz}{\vec z}

\nc{\vB}{\vec B}
\nc{\vC}{\vec C}
\nc{\vD}{\vec D}
\nc{\vE}{\vec E}
\nc{\vF}{\vec F}
\nc{\vG}{\vec G}
\nc{\vH}{\vec H}
\nc{\vP}{\vec P}
\nc{\vQ}{\vec Q}
\nc{\vR}{\vec R}
\nc{\vS}{\vec S}
\nc{\vV}{\vec V}
\nc{\vW}{\vec W}
\nc{\vX}{\vec X}
\nc{\vY}{\vec Y}
\nc{\vZ}{\vec Z}

\nc{\ol}{\overline}
\nc{\abar}{\ol{a}}
\nc{\bbar}{\ol{b}}
\nc{\cbar}{\ol{c}}
\nc{\dbar}{\ol{d}}
\nc{\ebar}{\ol{e}}
\nc{\fbar}{\ol{f}}
\nc{\gbar}{\ol{g}}
\nc{\ibar}{\ol{\imath}}
\nc{\jbar}{\ol{\jmath}}
\nc{\kbar}{\ol{k}}
\nc{\lbar}{\ol{l}}
\nc{\mbar}{\ol{m}}
\nc{\nbar}{\ol{n}}
\nc{\pbar}{\ol{p}}
\nc{\qbar}{\ol{q}}
\nc{\rbar}{\ol{r}}
\nc{\sbar}{\ol{s}}
\nc{\ubar}{\ol{u}}
\nc{\vbar}{\ol{v}}
\nc{\wbar}{\ol{w}}
\nc{\xbar}{\ol{x}}
\nc{\ybar}{\ol{y}}
\nc{\zbar}{\ol{z}}

\nc{\Abar}{\ol{A}}
\nc{\Bbar}{\ol{B}}
\nc{\cBbar}{\ol{\cB}}
\nc{\Cbar}{\ol{C}}
\nc{\Dbar}{\ol{D}}
\nc{\Ebar}{\ol{E}}
\nc{\Fbar}{\ol{F}}
\nc{\Gbar}{\ol{G}}
\nc{\Jbar}{\ol{J}}
\nc{\Kbar}{\ol{K}}
\nc{\cKbar}{\ol{\cK}}
\nc{\Lbar}{\ol{L}}
\nc{\cLbar}{\ol{\cL}}
\nc{\Mbar}{\ol{M}}
\nc{\Nbar}{\ol{N}}
\nc{\Pbar}{\ol{P}}
\nc{\Qbar}{\ol{Q}}
\nc{\Rbar}{\ol{R}}
\nc{\Sbar}{\ol{S}}
\nc{\Tbar}{\ol{T}}
\nc{\Ubar}{\ol{U}}
\nc{\Vbar}{\ol{V}}
\nc{\cVbar}{\ol{\cV}}
\nc{\Wbar}{\ol{W}}
\nc{\cWbar}{\ol{\cW}}
\nc{\Xbar}{{\overline X}}
\nc{\Ybar}{{\overline Y}}
\nc{\Zbar}{{\overline Z}}
\nc{\cZbar}{{\overline \cZ}}

\nc{\epsbar}{\ol{\epsilon}}
\nc{\albar}{\ol{\al}}
\nc{\Albar}{\ol{\Al}}
\nc{\betabar}{\ol{\beta}}
\nc{\Betabar}{\ol{\Beta}}
\nc{\lambar}{\ol{\lambda}}
\nc{\kapbar}{\ol{\kappa}}
\nc{\zetabar}{\ol{\zeta}}
\nc{\Zetabar}{\ol{\Zeta}}
\nc{\taubar}{\ol{\tau}}
\nc{\Taubar}{\ol{\Tau}}
\nc{\psibar}{\ol{\psi}}
\nc{\Psibar}{\ol{\Psi}}
\nc{\tpsibar}{\ol{\tpsi}}
\nc{\tPsibar}{\ol{\tPsi}}
\nc{\phibar}{\ol{\phi}}
\nc{\Phibar}{\ol{\Phi}}
\nc{\chibar}{\ol{\chi}}
\nc{\mubar}{\ol{\mu}}
\nc{\nubar}{\ol{\nu}}
\nc{\rhobar}{\ol{\rho}}
\nc{\ombar}{\ol{\om}}
\nc{\Ombar}{\ol{\Om}}
\nc{\Deltabar}{\ol{\Delta}}
\nc{\Thetabar}{\ol{\Theta}}
\nc{\xibar}{\ol{\xi}}
\nc{\Xibar}{\ol{\Xi}}

\nc{\Dthbar}{\ol{\rm D3}}

\nc{\fdot}{\dot{f}}
\nc{\gdot}{\dot{g}}
\nc{\pdot}{\dot{p}}
\nc{\qdot}{\dot{q}}
\nc{\rdot}{\dot{r}}
\nc{\sdot}{\dot{s}}
\nc{\tdot}{\dot{t}}
\nc{\udot}{\dot{u}}
\nc{\vdot}{\dot{v}}
\nc{\wdot}{\dot{w}}
\nc{\xdot}{\dot{x}}
\nc{\xddot}{\ddot{x}}
\nc{\ydot}{\dot{y}}
\nc{\zdot}{\dot{z}}
\nc{\yddot}{\ddot{y}}

\nc{\Adot}{\dot{A}}
\nc{\Bdot}{\dot{B}}
\nc{\Cdot}{\dot{C}}
\nc{\Udot}{\dot{U}}
\nc{\Vdot}{\dot{V}}
\nc{\Wdot}{\dot{W}}

\nc{\taudot}{\dot{\tau}}
\nc{\phidot}{\dot{\phi}}
\nc{\psidot}{\dot{\psi}}
\nc{\chidot}{\dot{\chi}}
\nc{\sinp}{s_{\phi}}
\nc{\cosp}{c_{\phi}}
\nc{\tanp}{t_{\phi}}
\nc{\spone}{s_{\phi_1}}
\nc{\cpone}{c_{\phi_1}}
\nc{\tpone}{t_{\phi_1}}
\nc{\sptwo}{s_{\phi_2}}
\nc{\cptwo}{c_{\phi_2}}
\nc{\tptwo}{t_{\phi_2}}
\nc{\spth}{s_{\phi_3}}
\nc{\cpth}{c_{\phi_3}}
\nc{\tpth}{t_{\phi_3}}
\nc{\calp}{c_{\al}}
\nc{\salp}{s_{\al}}

\nc{\csch}{{\rm csch}}
\nc{\sech}{{\rm sech}}

\nc{\cothzlami}{\coth(z-\lam_i)}
\nc{\coshzlami}{\cosh(z-\lam_i)}
\nc{\sinhzlami}{\sinh(z-\lam_i)}

\nc{\cothzlamj}{\coth(z-\lam_j)}
\nc{\coshzlamj}{\cosh(z-\lam_j)}
\nc{\sinhzlamj}{\sinh(z-\lam_j)}

\nc{\cothlamij}{\coth(\lam_i-\lam_j)}
\nc{\coshlamij}{\cosh(\lam_i-\lam_j)}
\nc{\sinhlamij}{\sinh(\lam_i-\lam_j)}

\nc{\bah}{{\mathbf {\hat{A}}}}
\nc{\bX}{{\mathbf X}}
\nc{\ba}{{\bf a}}
\nc{\bb}{{\bf b}}
\nc{\bc}{{\bf c}}
\nc{\bd}{{\bf d}}
\nc{\bg}{{\bf g}}
\nc{\bk}{{\bf k}}
\nc{\bl}{{\bf l}}
\nc{\bm}{{\bf m}}
\nc{\bn}{{\bf n}}
\nc{\bo}{{\bf o}}
\nc{\bp}{{\bf p}}
\nc{\bq}{{\bf q}}
\nc{\br}{{\bf r}}
\nc{\bs}{{\bf s}}
\nc{\bt}{{\bf t}}
\nc{\bu}{{\bf u}}
\nc{\bv}{{\bf v}}
\nc{\bw}{{\bf w}}
\nc{\bx}{{\bf x}}
\nc{\by}{{\bf y}}
\nc{\bz}{{\bf z}}
\nc{\bom}{{\bf \om}}
\nc{\bombar}{{\mathbf \ombar}}
\nc{\bPhi}{{\bf \Phi}}

\nc{\rma}{{\rm a}}
\nc{\rmb}{{\rm b}}
\nc{\rmc}{{\rm c}}
\nc{\rmd}{{\rm d}}
\nc{\rmg}{{\rm g}}
\nc{\rk}{{\rm k}}
\nc{\rml}{{\rm l}}
\nc{\rmm}{{\rm m}}
\nc{\rmn}{{\rm n}}
\nc{\rmo}{{\rm o}}
\nc{\rmp}{{\rm p}}
\nc{\rmq}{{\rm q}}
\nc{\rmr}{{\rm r}}
\nc{\rms}{{\rm s}}
\nc{\rmt}{{\rm t}}
\nc{\rmu}{{\rm u}}
\nc{\rmv}{{\rm v}}
\nc{\rmw}{{\rm w}}
\nc{\rmx}{{\rm x}}
\nc{\rmy}{{\rm y}}
\nc{\rmz}{{\rm z}}

\nc{\dal}{\dot{\al}}
\nc{\thadot}{\dot{\tha}}
\nc{\thab}{\bar{\theta}}
\nc{\thal}{\theta^{\al}}
\nc{\thdal}{\bar{\theta}^{\dal}}

\nc{\thsigthm}{\tha \sigma^m \thab}
\nc{\thsigthn}{\tha \sigma^n \thab}

\nc{\Dal}{D_{\al}}
\nc{\Ddal}{\bar{D}_{\dal}}
\nc{\CDal}{{\cal D}_{\al}}
\nc{\CDdal}{\bar{\cal D}_{\dal}}

\nc{\eq}[1]{{(\ref{#1})}}
\nc{\eqtwo}[2]{{(\ref{#1},\ref{#2})}}
\nc{\eqthree}[3]{(\ref{#1},\ref{#2},\ref{#3})}
\nc{\eqfour}[4]{(\ref{#1},\ref{#2},\ref{#3},\ref{#4})}
\nc{\eqfive}[5]{(\ref{#1},\ref{#2},\ref{#3},\ref{#4,\ref{#5}})}
\nc{\non}{\nonumber}
\nc{\Fzero}{F_{(0)}}
\nc{\Ftwo}{F_{(2)}}
\nc{\Ffour}{F_{(4)}}
\nc{\Fone}{F_{(1)}}
\nc{\Fthree}{F_{(3)}}
\nc{\Ffive}{F_{(5)}}
\nc{\Fn}{F_{(n)}}
\nc{\Fp}{F_{(p)}}

\nc{\tFzero}{\tF_{(0)}}
\nc{\tFtwo}{\tF_{(2)}}
\nc{\tFfour}{\tF_{(4)}}
\nc{\tFone}{\tF_{(1)}}
\nc{\tFthree}{\tF_{(3)}}
\nc{\tFfive}{\tF_{(5)}}
\nc{\tFn}{\tF_{(n)}}
\nc{\tFp}{\tF_{(p)}}

\nc{\Czero}{C_{(0)}}
\nc{\Ctwo}{C_{(2)}}
\nc{\Cfour}{C_{(4)}}
\nc{\Cone}{C_{(1)}}
\nc{\Cthree}{C_{(3)}}
\nc{\Cfive}{C_{(5)}}
\nc{\Cn}{C_{(n)}}

\nc{\equ}{{\rm eq}}
\def\Im{{\rm Im \hspace{0.5mm} }}

\def\Re{{\rm Re \hspace{0.5mm}}}

\nc{\vol}{{\rm vol}}
\nc{\Ainf}{A_{\infty}}
\nc{\End}{{\rm End}}
\nc{\Ext}{{\rm Ext}}
\nc{\IIB}{{\rm IIB}}
\nc{\Ad}{{\rm Ad}}
\nc{\IIA}{{\rm IIA}}
\nc{\AdS}{{\rm AdS}}
\nc{\CFT}{{\rm CFT}}
\nc{\diag}{{\rm diag}}
\nc{\Log}{{\rm Log}}
\nc{\Dslash}{\ensuremath \raisebox{0.025cm}{\slash}\hspace{-0.32cm} D}
\nc{\cDslash}{\ensuremath \raisebox{0.025cm}{\slash}\hspace{-0.32cm} \cD}
\nc{\omslash}{\om\!\!\!/}
\nc{\no}{\!:\!\!}
\nc{\ointdz}{\oint\frac{dz}{2\pi i}}
\nc{\ointdzone}{\oint\frac{dz_1}{2\pi i}}
\nc{\ointdztwo}{\oint\frac{dz_2}{2\pi i}}
\nc{\ointdzb}{\oint\frac{d\zbar}{2\pi i}}
\nc{\ointdzbone}{\oint\frac{d\zbar_1}{2\pi i}}
\nc{\ointdzbtwo}{\oint\frac{d\zbar_2}{2\pi i}}
\nc{\dz}{\frac{dz}{2\pi i}}
\nc{\dzb}{\frac{d\zbar}{2\pi i}}
\nc{\bpm}{\begin{pmatrix}}
\nc{\epm}{\end{pmatrix}}
 \nc{\bitem}{\begin{itemize}}
 \nc{\eitem}{\end{itemize}}
 \nc{\exercise}{\vskip 2mm \noindent {\bf Exercise:}}
 \nc{\definition}{\vskip 2mm \noindent {\bf Definition:}}
 
\begin{document}
%%%%%%%%%%%%%%%%%%%%%%%%%%%%%%%%%%%%%%%%%%%%%%%%%%%%%%%%%%%%%%%%%
%%%%%%%%%%%%%%%%%%%%%%%%%%%%%%%%%%%%%%%%%%%%%
\begin{center}
\end{center}

\vspace{0.5cm}
\begin{center}
\baselineskip=13pt {\LARGE \bf{
Abelian Hypermultiplet Gaugings \\ and BPS Vacua in $\cN=2$ Supergravity }}
 \vskip1.5cm 
Harold Erbin and Nick Halmagyi\\ 
\vskip0.5cm
\textit{Sorbonne Universit\'es, UPMC Paris 06,  \\ 
UMR 7589, LPTHE, 75005, Paris, France \\
and \\
CNRS, UMR 7589, LPTHE, 75005, Paris, France}\\
\vskip0.5cm
harold.erbin@lpthe.jussieu.fr \\ 
halmagyi@lpthe.jussieu.fr \\ 

\end{center}

\begin{abstract}
We analyze the gauging of Abelian isometries on the hypermultiplet scalar manifolds of $\cN=2$ supergravity in four dimensions. This involves a study of symmetric special quaternionic-K\"ahler manifolds, building on the work of de Wit and Van Proeyen. In particular we compute the general set of Killing prepotentials and associated compensators for these manifolds. This allows us to glean new insights about AdS$_4$ vacua which preserve the full $\cN=2$ supersymmetry as well as BPS static black hole horizons.
\end{abstract}
\newpage
\tableofcontents

%%%%%%%%%%%%%%%%%%%%%%%%%%%%%%%%%%%%
\section{Introduction}
%%%%%%%%%%%%%%%%%%%%%%%%%%%%%%%%%%%%

There has been much work in the last 10 years deriving gauged supergravity theories in four dimensions from string theory and M-theory. Such theories have applications to string phenomenology, holography and black hole physics. The canonical vacua in $\cN=2$ gauged supergravity are the $\cN=2$ AdS$_4$ vacua; the equations for such vacua are straightforward to derive, one can find recent discussions in \cite{Hristov:2009uj, Louis:2012ux, deWit:2011gk} which we build upon in the current work. 

The essential step in gauging supergravity theories is to charge the gravitino  under isometries of the vector scalar manifold $\cM_v$ and the hypermultiplet scalar manifold $\cM_h$. We will consider Abelian gaugings in this work so that the only other charged fields are the hypermultiplets. Certainly one should understand these global symmetries before making them local and as luck would have it, the symmetries of very special K\"ahler manifolds and quaternionic K\"ahler manifolds in the image of a c-map have been studied in great depth some time ago by de Wit and Van Proeyen \cite{deWit:1990na, deWit:1992wf, deWit:1993rr}. In the current work we utilize these descriptions of symmetric quaternionic K\"ahler manifolds to compute the Killing prepotentials, a key ingredient in constructing gauged supergravity theories.

An interesting feature of quaternionic K\"ahler manifolds is that the curvature forms need not be exactly invariant under a given Killing vector but may transform under the $SU(2)$ holonomy group. We find that in order to have $\cN=2$ AdS$_4$ vacua, one must gauge along an isometry $k$ which induces such an $SU(2)$ transformation. In the language of the text below, this implies there is a non-trivial compensator $W^x_k$. Notably, the vector fields which generate the Heisenberg algebra do not generate such transformations and are thus not sufficient for the existence of $\cN=2$ AdS$_4$ vacua. Nonetheless there are large numbers of $\cN=1$ AdS$_4$ vacua found by gauging the Heisenberg algebra \cite{Cassani:2009ck, Cassani:2009na}.

We also analyze the conditions for quarter BPS black hole horizons of the form AdS$_2\times \Sig_g$ where $\Sig_g$ is a Riemann surface of genus $g$. For the same vacua but in FI-gauged supergravity, the algebraic BPS equations have been solved \cite{Halmagyi:2013qoa} and the entropy found to be related to the famous quartic invariant. We repeat this analysis for Abelian gaugings of hypermultiplets, and again find that the quartic invariant plays a prominent role. 

The work of \cite{deWit:1990na, deWit:1992wf, deWit:1993rr} considered homogeneous quaternionic manifolds which lie in the image of a c-map. This later condition is tantamount to the fact that they arise in three dimensions as the moduli space of vector in a dimensional reduction from four dimensions. In particular the symmetries of such manifolds were classified. We build on this work while just considering the symmetric quaternionic K\"ahler manifolds. While all homogeneous spaces are cosets, the condition of being symmetric means that all possible symmetries are realized and they form a semi-simple Lie algebra. We add a conceptual point to the analysis of  \cite{deWit:1990na, deWit:1992wf, deWit:1993rr}; the so-called {\it hidden} isometries must act symplectically on the base special K\"ahler manifold, this is not at all evident from the formulae of de Wit and Van Proeyen. By restricting to symmetric spaces we are able to demonstrate this explicitly although generalizing this to the homogeneous case is an interesting future step. 

This paper is organized as follows. In section 2  we review aspects of $\cN=2$ AdS$_4$ vacua as well as quarter BPS black hole horizons. In section 3 we review aspects of special K\"ahler geometry which we will need. In section 4 we present the symmetries of symmetric quaternionic K\"ahler manifolds. In section 5 we compute the prepotentials and compensators for all symmetries on these quaternionic K\"ahler manifolds. In section 6 we discuss the constraints on the embedding tensor from locality and in section 7 we discuss two examples from M-theory which utilize $\cM_h=G_{2(2)}/SO(4)$.

\vskip 5mm
\noindent {\bf Note added}: As this paper was being prepared for submission, we were made aware of a recent article \cite{Fre:2014pca} which overlaps with our work. In particular they also compute the Killing prepotentials associated to symmetric quaternionic K\"ahler manifolds.
%%%%%%%%%%%%%%%%%%%%%%%%%%%%%%%%%%%%
\section{BPS vacua in \texorpdfstring{$\cN=2$}{N=2} Gauged Supergravity}
%%%%%%%%%%%%%%%%%%%%%%%%%%%%%%%%%%%%

In this section we review some basic facts about gauged $\cN=2$ supergravity with $n_v$-vector multiplets and $n_h$-hypermultiplets. We then discuss the conditions for AdS$_4$ vacua with eight supercharges and AdS$_2\times \Sig_g$ vacua which preserve four supercharges.

The scalar kinetic terms respect a division into hyper-scalars $\{ q^u| u=1\,,\ldots , 4n_h\}$ and vector-scalars $\{ \tau^j = x^j +i y^j | j=1\,,\ldots, n_v  \}$:
\be
\cM_{{\rm scalar}}= \cM_v \times \cM_h
\ee
where $\cM_v$ is a special K\"ahler manifold and $\cM_h$ is a quaternionic K\"ahler manifold. The gauging procedure involves minimally coupling certain scalar fields with respect to a chosen set of isometries of $\cM_{{\rm scalar}}$ and in this article we will exclusively consider gauging Abelian isometries on $\cM_h$. Accordingly, the hyper-scalars appear in the action with the covariant derivative
\be
Dq^u = dq^u + k^u_\Lam A^\Lam
\ee
where $\{A^\Lam| \Lam=0\,,\ldots , n_v\}$ are the vectors fields including the graviphoton. For each $\Lam$, the vector field $k_\Lam=k^u_\Lam \del_u$ on $\cM_h$ is Killing, one can thus associate to it a Killing prepotential  $P^x$:
\be
k_\Lam \lrcorner \Om^x = -D P^x_\Lam\,,
\ee
where $\Om^x$ is the triplet of curvature two-forms as described in appendix \ref{app:quaternionic}. Much as the Killing vector provides the charge for the hyper-scalars, this Killing prepotential provides the charge for the doublet of gravitinos  $\Psi^A$:
\be
D \Psi^A_\mu = d \Psi^A_\mu + P^x_\Lam A^\Lam (\sig^x \eps)^A_{\ B} \Psi^B_\mu
\ee

Our metric ansatz is
\be
\label{eq:metric-ansatz}
ds_4^2 = -e^{2U}dt^2 + e^{-2U} dr^2 + e^{2(V-U)} d\Sig_g^2\,.
\ee
$d\Sigma_g^2$ is the uniform metric on Riemann surfaces
\begin{equation}
	\Sigma_g =
	\begin{cases}
		S^2 & \kappa = 1 \\
		T^2 & \kappa = 0 \\
		\mathbb H^2 / \tGam & \kappa = -1
	\end{cases}
\end{equation} 
where $\kappa$ is the curvature of $\Sigma_g$ and $\tGam$ is a Fuchsian group which do not enter in our local analysis.

On $AdS_4$ and $AdS_2 \times \Sigma_g$ vacua these functions $U$ and $V$ are respectively
\begin{align}
	AdS_4 :& \qquad e^U = \frac{r}{R}, \qquad e^V = \frac{r^2}{R}, \\
	AdS_2 \times \Sigma_g :& \qquad e^U = \frac{r}{R_1}, \qquad e^V = \frac{R_2}{R_1}\, r. \label{AdS2Sigg}
\end{align}

The gauged fields give rise to the charges
\bea
p^\Lam &=&\frac{1}{4\pi} \int_{\Sig_g} F^\Lam\,,\quad\quad q_\Lam =\frac{1}{4\pi} \int_{\Sig_g} G_\Lam 
\eea
where $F^\Lambda = d A^\Lambda$ and the dual field strength is
\be
G_\Lam= \cR_{\Lam \Sig} F^\Sig - \cI_{\Lam \Sig} * F^\Sig\,,
\ee
the matrices $\cR$ and $\cI$ being defined in the appendix~\ref{sec:special-geometry}.

%%%%%%%%%%%%%%%%%%%%%%%%%%%%%%%%%%%%
\subsection{Magnetic Gaugings}\label{sec:magnetic-gaugings}
%%%%%%%%%%%%%%%%%%%%%%%%%%%%%%%%%%%%
A key step in the development of gauged supergravity is making symmetries local with respect to magnetic gauge fields in addition to the more canonical electric gauge fields. In general this can be neatly formulated in terms of the embedding tensor \cite{deWit:2005ub, Samtleben:2008pe} but since we will be restricting to Abelian gaugings we find it clearer to merely include the magnetic Killing vectors $\tk_\Lam^u$ and magnetic Killing prepotentials $\tP^x_\Lam$, we use the following notation for the symplectic vector of gauging parameters
\be
\cK^u = \bpm \tk^{u,\Lam} \\ k^u_\Lam \epm\,,\quad\quad \cP^x= \bpm \tP^{x,\Lam} \\ P^x_\Lam \epm \,.
\ee

In section \ref{sec:gaugingconstraints} we will enforce a particular set of constraints on these objects to ensure that there exists a symplectic frame where all the gaugings are electric \cite{deWit:2005ub}. If one is willing to consider an arbitrary prepotential $\cF$ one could thus equally well consider solely electric gaugings from the outset but we will allow for magnetic gaugings and restrict the class of prepotentials $\cF$ which we consider.

%%%%%%%%%%%%%%%%%%%%%%%%%%%%%%%%%%%%
\subsection{\texorpdfstring{$\cN=2$ AdS$_4$}{N=2 AdS4} Equations}\label{sec:AdS4Vac-BPS}
%%%%%%%%%%%%%%%%%%%%%%%%%%%%%%%%%%%%

We first  analyze the algebraic equations for $\cN=2$ AdS$_4$ vacua with radius $R$ and constant scalar fields:
\bea
\langle \cP^x , D_i \cV \rangle &=& 0\label{AdSEq1}\\
\cL^x \cLbar^x&=& \frac{1}{R^2} \label{AdSEq2}\\
 \langle  \cK^u, \cV\rangle &=& 0\,. \label{KVeq}
 \eea
 where we have used the fairly standard definition
 \be
 \cL^x=\langle \cP^x,\cV\rangle\,.
 \ee
To simplify these equations somewhat we first perform a symplectic rotation to a frame where $\cP^x$ is purely electric (i.e. $\tP^x_\Lam=0$), then \eq{AdSEq1} reduces 
\be
P^x_\Lam f_i^\Lam=0\,. \label{Pf}
\ee
This implies that for each $x$, $P^x_\Lam$ is orthogonal to $f_i^\Lam$ for each $i$ and thus
\be
P^x_\Lam=c^x P_\Lam \label{PcP}
\ee 
for some functions $c^x(q^u)$. A local $SU(2)$-transformation which can be used to set 
\be
c^1=c^2=0\,.
\ee
We emphasize that \eq{PcP} must be {\it enforced} for by solving \eq{Pf}, it is not a generic consequence of the theory.

As a result \eq{AdSEq1}-\eq{KVeq} become
\bea
\cP&=& -2\Im \bslb \cLbar \cV \bsrb \label{AdSEq3}\\
\cL&=& \frac{ie^{i\psi}}{R} \label{AdSEq4}\\
 \langle  \cK^u, \cV\rangle &=& 0 \label{KVeq2}
 \eea
 where we have introduced
 \be
 \cP\equiv \cP^3\,,\quad\quad \cL\equiv \cL^3\,.
 \ee
In this work, our strategy to solve these equations will be to first recognize that \eq{AdSEq3} and \eq{AdSEq4} are identical in form to the AdS$_4$ equations in FI-gauged supergravity \cite{Gnecchi:2013mta}, which are in turn identical in form to the attractor equations in {\it ungauged} $\cN=2$ supergravity \cite{Ferrara:1995ih, Ferrara:1996dd} and can be solved quite explicitly \cite{Shmakova:1996nz}. When $\cM_v$ is a symmetric space as well as a very special K\"ahler manifold, we can use the identity \eq{ReVImV} and \eq{AdSEq3} to transform \eq{KVeq2} into
\bea
0&=&I_4(\cK^u,\cP,\cP,\cP)\sim  \nabla^u  I_4(\cP)  
\eea
which is $4n_h$ equations depending only on the hypermultiplet scalars.

One objective of our current work is to clarify \eq{KVeq2} and to do so we first recall an argument from \cite{Galicki:1987, D'Auria:1990fj} regarding $SU(2)$ compensating transformations. For a given Killing vector $k$, the spin connection on $\cM_h$ need only be invariant under the Lie derivative by $k$ up to a gauge transformation
\be
\cL_k \om^x= \nabla W^x_k\,.
\ee
Using this one can algebraically relate the Killing prepotential associated to $k$
\be
P^x_k= k \lrcorner \om^x-W^x_k\,.
\ee
Simple inspection of \eq{KVeq2} shows that if none of the gauged isometries of $\cM_h$ have a non-trivial compensator then $\cL=0$ which by \eq{AdSEq4}  does not give a regular AdS$_4$ vacuum.  We see that a necessary condition in order to have a regular $\cN=2$, AdS$_4$ vacuum is that one must gauge along at least one isometry of $\cM_h$ which has a non-trivial compensator $W^x_\Lam$ and much of this paper is devoted to fleshing out this idea in some detail. We will build upon the work of Van Proeyen and de Wit \cite{deWit:1990na, deWit:1992wf} where they classified isometries of particular quaternionic-K\"ahler manifolds but we will provide simplified formulae for these isometries which we consider more easily utilized in gauged supergravity, in particular we compute the compensators $W^x_\Lam$.

 %%%%%%%%%%%%%%%%%%%%%%%%%%%%%%%%%%%%
\subsection{BPS Black Hole Horizons: \texorpdfstring{{\rm AdS}$_2\times \Sig_g$}{AdS2 x Sigma}} \label{sec:BlackHorizons}
%%%%%%%%%%%%%%%%%%%%%%%%%%%%%%%%%%%%

Another canonical vacuum in four dimensional $\cN=2$ gauged supergravity is AdS$_2\times \Sig_g$. The bosonic fields and the supersymmetry parameter are independent of the co-ordinates on $\Sig_g$ which thus allows for quotient of $\HH^2$ by $\tGam$. We refer to such solutions as {\it black hole horizons} since the horizon of a static extremal black hole is of this form. The solutions which we study of this form preserve two real Poincar\'e supercharges plus two superconformal supercharges, they are typically referred to as quarter-BPS.

The equations for BPS black hole horizons with hypermultiplets were derived in \cite{Halmagyi:2013sla}.  We will use the symplectic completion of these equations but as explained above, once the locality constraints of section \ref{sec:gaugingconstraints} are imposed, these models can always be symplectically rotated to a frame with purely electric gaugings at the cost of a potentially non-trivial transformation on the prepotential. From the equations in appendix \ref{sec:appendix-BH-eqs} we note that the Killing prepotentials always appear in terms of the quantity $P^x_p\equiv P^x_\Lam p^\Lam$. Since by \eq{pprime} $p^\Lam$ are constant, we can use a local (on $\cM_h$) $SU(2)$ transformation to set 
\be
P^1_p=P^2_p=0.
\ee
In this way, much like the AdS$_4$ equations, the BPS equations depend only on $P^3_\Lam$.

The BPS equations for AdS$_2\times \Sig_g$ solutions are 
\bea
\cQ - R_2^2 \cM \cP&=& -4\, \Im (\cZbar\cV)\label{AdS2Sigeq1}\\
\cZ&=& e^{i\psi}\frac{R_2^2}{2R_1} \label{AdS2Sigeq2}\\
\langle \cP , \cQ \rangle &=& \kappa \label{AdS2Sigeq3}\\
\langle\cK^u,\cV \rangle&=& 0 \label{KVzero}\\
\langle\cK^u,\cQ \rangle&=& 0\label{AdS2Sigeq5}
\eea
where $(R_1,R_2)$ are the radii of AdS$_2$ and $\Sig_g$ in the metric ansatz \eq{eq:metric-ansatz} and \eq{AdS2Sigg}. When $\cM_v$ is a symmetric space, the equations \eq{AdS2Sigeq1}-\eq{AdS2Sigeq3} were explicitly solved in \cite{Halmagyi:2013qoa} and implicitly solved when $\cM_v$ is not symmetric. This solution can then be used to reduce the full set \eq{AdS2Sigeq1}-\eq{AdS2Sigeq5} to \eq{KVzero}-\eq{AdS2Sigeq5} depending only on the hypermultiplet scalars $q^u$. Of course these remaining equations depend non-trivially on the gauging parameters. Using the results of \cite{Halmagyi:2014qza} one can replace $\cV$ in \eq{KVzero} with an expression involving $I'_4$ evaluated on $\cP$ and $\cQ$.

We note that as in \cite{Halmagyi:2013qoa} the entropy of the black hole is obtained by expanding
\be
0=I_4\blp \cQ-i R_2^2 \cP\brp
\ee
into real and imaginary parts
\be
R_2^4=\frac{-I_4(\cQ,\cQ,\cP,\cP)\pm \sqrt{I_4(\cQ,\cQ,\cP,\cP)^2-I_4(\cQ,\cQ,\cQ,\cQ)I_4(\cP,\cP,\cP,\cP)}}{I_4(\cP,\cP,\cP,\cP)}\,.
\ee 
We note that the $\cP$ depends nontrivially on $q^u$ (as opposed to the constant gauge couplings in the FI-gauged supergravity studied in \cite{Halmagyi:2013qoa}) which must in turn be evaluated by solving \eq{KVzero} and \eq{AdS2Sigeq5}.

 %%%%%%%%%%%%%%%%%%%%%%%%%%%%%%%%%%%%
\section{Symmetries of Special K\"ahler Manifolds}\label{SymmKahler}
%%%%%%%%%%%%%%%%%%%%%%%%%%%%%%%%%%%%

We warm up by recalling various features of the symmetry structure of special K\"ahler manifolds \cite{deWit:1990na, deWit:1992wf}. In general for homogeneous spaces, there are certain universal symmetries which are guaranteed to exist for any such manifold and then there are model dependent symmetries which are constrained. For symmetric spaces all the model dependent symmetries are realized. This is particularly useful for our computations in the next section where the so-called {\it hidden} isometries act symplectically on the base special K\"ahler manifold.

A key point regarding symmetries on special K\"ahler manifolds is that all symmetries act on the symplectic sections as linear symplectic transformations:
\bea
\delta \bpm  X^\Lam \\ F_\Lam \epm&=&\UU \bpm  X^\Lam \\ F_\Lam \epm\,, \label{SymplAction} 
\eea
with
\be
\UU=\bpm \cQ & \cR \\ \cS & \cT\epm\,, \quad\quad\quad \cR=\cR^T\,,\quad\quad \cS=\cS^T\,,\quad\quad \cT=-\cQ^T \,. \label{UUcomponents}
\ee
However not all symplectic transformations generate isometries of $\cM_v$, true symmetries are constrained by
\be
\delta F_\Lam = \frac{\del F_\Lam }{\del X^\Sig}\, \delta X^\Sig\,,
\ee
contracting both sides and using the homogeneity of $F_\Lam$ we get
\bea
X^\Lam \delta F_\Lam &=& F_\Lam  \delta X^\Sig \quad\quad\Rightarrow \quad\quad 0= X^\Lam  \cS_{\Lam \Sig} X^\Sig  - 2 X^\Lam (\cQ^T)_\Lam^{\ \Sig} F_\Sig - F_\Lam \cR^{\Lam \Sig}F_{\Sig}\,. \label{CubicConstraint}
\eea
This constraint is sufficient to classify isometries on special K\"ahler manifolds.

%%%%%%%%%%%%%%%%%%%%%%%%%%%%%%%%%%%%
\subsection{Cubic Prepotentials}
%%%%%%%%%%%%%%%%%%%%%%%%%%%%%%%%%%%%
When the prepotential is cubic
\be
\cF=-d_{ijk}\frac{X^i X^j X^k}{X^0}\,,
\ee
the general solution to \eq{CubicConstraint} is found by expanding in powers of $\tau^i$ and one finds
\be
\cQ^{\Lam}_{\ \Sig}= -(\cT^T)^{\Lam}_{\ \Sig} = \bpm \beta& a_j \\ b^i & B^{i}_{\ j}+ \frac{1}{3} \beta \delta^i_{\ j} \epm,\ \ \ \ 
\cS_{\Lam\Sig}= \bpm 0 & 0 \\ 0 & -6d_{ijk} b^k \epm,\ \ \ \
\cR^{\Lam \Sig}=  \bpm 0 & 0 \\ 0 & -\frac{3}{32} \hd^{ijk}a_k \epm  \label{QRSTdef}
\ee
where $\{\beta,B^i_{\ j},b^i,a_j\}$ are constants. On special coordinates these symmetries act as a generalization of fractional linear transformations:
\be
\delta \tau^i = b^ i -\frac{2}{3}\beta \tau^i + B^{i}_{\ j} \tau^j - \half R^{i\ \ \, l}_{\ jk}\tau^j \tau^k  a_l\,. \label{KspCubic}
\ee
The unconstrained symmetries are given by axion shifts generated by $b_i$ and a common rescaling generated by $\beta$. The other rescalings generated by $B^i_{\ j}$ are constrained
\be
0= d_{i(kl}B^i_{\ j)}
 \ee
as are the non-linear symmetries generated by $a_i$ which must satisfy
 \be
a_i E^i_{jklm}=0\,,
\ee
where
\bea
E^i_{jklm}&=& \hd^{ijk} d_{j(lm}d_{np)k}-\frac{64}{27}  \delta^{i}_{(m}d_{npl)}\,,\\
\hd^{ijk}&=& \frac{g^{il} g^{jm}g^{kn}d_{lmn}}{d_y^2}\,.
\eea

When $\cM_v$ is a symmetric space, $\hd^{ijk}$ has constant entries and $E^i_{jklm}=0$. 
Then the symmetry group of $\cM_v$ will be a simple Lie algebra where $b^i$ generate the lowering operators, $a_i$ generate  raising operators and $(\beta,B^i_{\ j})$ generate Cartan elements. We can use the constant tensor $\hd^{ijk}$ to define the quartic invariant
\bea
 \label{quarticdef}
 \cI_4(\cQ) &=&  -(q_\Lam p^\Lam)^2 + \frac{1}{16} p^0 \hd^{ijk} q_i q_j q_k - 4q_0 d_{ijk} p^i p^j p^k +\frac{9}{16} \hd^{ijk}d_{klm} p^l p^m q_i q_j,
\eea
one can check that $\cI_4(\cQ)$ is invariant under the action
\be
\delta \bpm p^\Lam \\ q_\Lam \epm = \UU \bpm p^\Lam \\ q_\Lam \epm 
\ee
with $\UU$ given by \eq{UUcomponents} and \eq{QRSTdef}.

%%%%%%%%%%%%%%%%%%%%%%%%%%%%%%%%%%%%
\subsection{Quadratic Prepotentials} 
%%%%%%%%%%%%%%%%%%%%%%%%%%%%%%%%%%%%
We will also consider the solution to \eq{CubicConstraint} for the series of special K\"ahler manifolds which arise from quadratic prepotentials: 
\be
\cF=X^\Lam \eta_{\Lam \Sig} X^\Sig\,.
\ee
As explained in appendix \ref{app:quadraticprep} where more details are given, one can in general take 
\be
\eta=\frac{1}{2i}\diag\{1,-1,\ldots ,-1\}
\ee 
and $\cM_v$ is the homogeneous space
\be
\cM_v= \frac{SU(1,n_v)}{U(1)\times SU(n_v)}\,.
\ee
The solution to \eq{CubicConstraint} is given by \eq{UUcomponents} with
\bea
\cS_{\Lam \Sig}&=&4\eta_{\Lam\Upsilon} \cR^{\Upsilon \Delta} \eta_{\Delta \Sig}\,,\label{quadSR} \\
\cQ^0_{\ i}&=&\cQ^i_{\ 0}\,,\quad\quad \cQ^i_{\ j} = -\cQ^{j}_{\ i}\quad\quad \cQ^{\Lam}_{\ \Lam}=0\,, \label{quadQ}
\eea
with no summation on $\Lam$ in the last line. The special coordinates $\tau^i$ transform as
\bea
\delta \tau^i&=&\cA^{i}_{\ 0}
+\blp \cA^{i}_{\ j} - \cA^{0}_{\ 0} \delta^i_j \brp \tau^j
- \tau^i \tau^j \cA^{0}_{\ j} \label{KspQuad}
\eea
where
\be
\cA= \cQ+ 2 \cR \eta\,.
\ee
There is a unique (up to constant rescalings) quadratic invariant, given by
\be
\cJ_2(p^\Lam,q_\Lam)=4 p^\Lam \eta_{\Lam \Sig}p^\Sig -q_\Lam (\eta^{-1})^{\Lam \Sig} q_{\Sig}
\ee
which gives rise to the unique quartic invariant
\be
\cJ_4(p^\Lam,q_\Lam)=\bslb \cJ_2(p^\Lam,q_\Lam)\bsrb ^2\,. \label{J4def}
\ee

%%%%%%%%%%%%%%%%%%%%%%%%%%%%%%%%%%%%
\subsection{Lie derivative of the K\"ahler potential}
%%%%%%%%%%%%%%%%%%%%%%%%%%%%%%%%%%%%
Despite the fact that in homogeneous coordinates the K\"ahler potential is manifestly symplectic invariant
\be
e^{-K}= -i X^T \Om \Xbar  \label{KPotdef}
\ee
and the Killing vectors act by a linear symplectic transformation \eq{SymplAction}, in special coordinates the K\"ahler potential need not be exactly invariant under the action of the Killing vectors. For any given Killing vector $k$, there may be a compensating K\"ahler transformation 
\be
\cL_kK=f_k+\fbar_k\,.
\ee

For cubic prepotentials we have
\be
e^{-K}=8d_y\,,\quad\quad
\cL_\UU(K)=  2 \beta + 2 a_i x^i
\ee
giving the holomorphic function
\be
f_\UU =\beta+a_i \tau^i\,.
\ee
For quadratic prepotentials the K\"ahler potential is
\bea
e^{-K}=2(-1+\sum_{i=1}^{n_v}|\tau^i|^2)
\eea
the Lie derivative induces the K\"ahler transformation
\bea
f_{\UU}(\tau^i)=2 \tau^i \ol{\cA}^i_{\ 0}\,.
\eea
In a conceptually similar vein, we will find below that the action of various symmetries on our quaternionic K\"ahler manifolds induce non-trivial $SU(2)$ compensating transformations.

%%%%%%%%%%%%%%%%%%%%%%%%%%%%%%%%%%%%
\section{Symmetries of Special Quaternionic K\"ahler Manifolds}
%%%%%%%%%%%%%%%%%%%%%%%%%%%%%%%%%%%%

Many of the symmetries on a  special quaternionic K\"ahler manifold are constructed from the symmetries of the base special K\"ahler manifold $\cM_z$ and this is the reason for reviewing such symmetries in section \ref{SymmKahler}.
As reviewed in appendix \ref{app:quaternionic} the metric on a special quaternionic K\"ahler manifold which lies in the image of a c-map is\footnote{We will sometimes use the coordinate $\rho = e^{-2\phi}$.}
\bea
ds_{QK}^2 = d\phi^2 + g_{a\bbar} dz^a d\zbar^{\bbar} +\frac{1}{4} e^{4\phi }\blp d\sig + \half \xi^T \CC d\xi\brp^2 -\frac{1}{4}e^{2\phi} d\xi^T \CC \MM d\xi \label{quartmet}
\eea
with $a = 1, \ldots n_h-1$.
The symmetries of such manifolds have been studied in \cite{deWit:1990na, deWit:1992wf} and here we find somewhat more compact expressions and compute the Killing prepotentials. We use the notation whereby the symplectic sections on the special K\"ahler base $\cM_z$ and the symplectic vector for the Heisenberg fiber are denoted
\bea
&&Z=\bpm  Z^A \\ G_A\epm\,,\quad\quad  \xi=\bpm \xi^A \\ \txi_A \epm, \qquad
A = 0, \ldots n_h - 1\,,\\
&& Z^A=\bpm 1 \\ z^a\epm\,,\quad\quad a=1,\ldots , n_h-1 
\eea
While de Wit and Van Proeyen considered special quaternionic K\"ahler manifolds which are homogeneous spaces, we will focus on the symmetric spaces for simplicity.

One conceptual addition we add to the work of \cite{deWit:1990na, deWit:1992wf} is the following. Since the quaternionic K\"ahler metric $ds_{QK}^2$ has terms quadratic but not linear in $z^a$, any Killing vector which acts on the $z^a$ must be a linear symplectic transformation on the sections $Z$ of the form $\UU$ described above.  This transformation may have components $\{a_c,b^c,\beta,B^a_{\ b}\}$ which depend on the fields $\{\phi,\sig,\xi^A,\txi_A\}$. In this section we present this transformation for symmetric spaces, leaving the more complicated homogeneous spaces for future work.

We have computed the Killing vectors presented in this section by explicit computation using \cite{deWit:1990na, deWit:1992wf} as a guide but altering and correcting their formulae where necessary.

%%%%%%%%%%%%%%%%%%%%%%%%%%%%%%%%%%%%
\subsection{The Duality Symmetries}
%%%%%%%%%%%%%%%%%%%%%%%%%%%%%%%%%%%%

The so-called {\it duality} symmetries are generated by 
\bea
h_{\eps_+}&=& \frac{\del}{\del \sig} \,, \\
h_{\al}  &=& \CC \Bslb \del_\xi +\frac{1}{2} \xi \frac{\del}{\del \sig} \Bsrb \,,\\ 
h_{\eps_0} &=&\frac{\del}{\del \phi}-2\sig\frac{\del}{\del \sig}+ \xi \CC \del_\xi \,, \label{Killphi} \\
h_{\UU}&=&(\UU Z)^{A} \frac{\del }{\del Z^{A}}+(\UU \Zbar)^{A} \frac{\del }{\del \Zbar^{A}} - (\UU \xi)^T \CC \del_\xi
\label{KillU}
\eea
where
\be
\del_\xi =\bpm\frac{\del}{\del \txi_A} \\ -\frac{\del}{\del \xi^A} \epm.
\ee

The Killing vector $h_{\eps_+}$ is an axion shift while $h_\al$ are shifts of the Heisenberg fibers embelished with a field dependent shift of $\sig$ (there are $2 n_h$ of them). The Killing vector $h_{\eps_0}$ generates a universal scaling symmetry. These symmetries are all model independent, they exist for any special quaternionic K\"ahler manifold.
 
The Killing vector $h_{\UU}$ uses the symplectic matrix from (\ref{SymplAction}, \ref{UUcomponents}) and should be understood as a general Killing vector of the base special K\"ahler manifold $\cM_z$ which has been uniquely lifted to a Killing vector on $\cM_h$ (parameters are written for this vector since there is no symplectic expression without writing them). For cubic prepotentials, such symmetries with non-trivial $(b^a,\beta)$ are therefore universal while those with non-trivial $(B^a_{\ b},a_a)$ are constrained with all the $a_a$ symmetries being realized when $\cM_z$ is a symmetric space. The series of quadratic prepotentials are all symmetric spaces and all the symmetries of the base $\cM_z$ extend to symmetries of $\cM_h$.

The Killing vectors $h_\alpha$ are  $2n_h$-dimensional and their components read explicitly
\begin{equation}
	h_A = - \frac{\del}{\del \xi^A} + \frac{1}{2}\, \txi_A\, \frac{\del}{\del \sigma} \,, \qquad
	h^A = - \frac{\del}{\del \tilde \xi_A} - \frac{1}{2}\, \xi^A\, \frac{\del}{\del \sigma} \,.
\end{equation} 

%%%%%%%%%%%%%%%%%%%%%%%%%%%%%%%%%%%%
\subsection{The Hidden Symmetries}\label{sec:HiddenSym}
%%%%%%%%%%%%%%%%%%%%%%%%%%%%%%%%%%%%

In addition to these duality symmetries there are the so-called {\it hidden} symmetries and these have a more formidable expression. After some lengthy but unenlightening computations we find the Killing vectors fields to be generated by
\bea
\label{eq:hidden-killing}
h_{\eps_-}&=& - \sig \frac{\del}{\del\phi} +(\sig^2-e^{-4\phi}-W)\frac{\del}{\del\sig} - \sig \xi \CC \del_\xi + (\del_\xi W)^T \CC \del_\xi + \left[ (\ul{S} Z)^A \frac{\del }{\del Z^{A}} + c.c. \right] \\ 
h_{\hal} &=& - \frac{1}{2} \CC \xi \frac{\del}{\del\phi} +\Bslb \frac{\sig}{2} \CC \xi+ \CC \del_\xi W \Bsrb \frac{\del}{\del\sig}  + \Bslb \sig 1\!\!1+ \frac{1}{2} \CC\xi\, \xi^T \CC +  \del_\xi  (\del_\xi W)^T  \CC  \Bsrb \del_\xi \non \\
&&- \left[ (\CC \del_\xi \ul{S}\, Z)^A \frac{\del }{\del Z^{A}} + c.c. \right]
\eea
where 
\be
W= \frac{1}{4} h(\xi^A,\txi_A) -\frac{1}{2} e^{-2\phi} \xi^T \CC \MM \xi
\ee
and $\underline{S}$ is the symplectic matrix:
\be
\underline{S}=\half  \Blp \xi \xi^T +\half   H\Brp \CC\,,\quad\quad
H= \bpm \del^I \del^J  h(\xi^A,\txi_A) & - \del^I \del_J h(\xi^A, \txi_A) \\ - \del_I \del^J h(\xi^A,\txi_A) & \del_I \del_J h(\xi^A, \txi_A) \epm = \del_\xi \, (\del_\xi h)^T \,.
\ee
In addition $h(\xi^A, \txi_A)$ is a particular quartic polynomial which will be elaborate on below. In appendix \ref{app:hidden-killing} we present these Killing vectors in a form more easily comparable with those in \cite{deWit:1990na, deWit:1992wf}, in fact we have amended an error in $h_{\eps_-}^\sig$ and $h_{\hal}^\sig$ which appears in those works.

One can schematically see that the general form of
\begin{equation}
	\label{deltaZG}
	\delta_{\epsilon_-} Z = \underline{S}\, Z, \qquad
	\delta_{\hal} Z = \CC \del_\xi \ul{S}\, Z
\end{equation} 
is necessary since the metric \eq{quartmet} has no terms linear in $dz^a$. As such when computing the variation of \eq{quartmet} with respect to any Killing vector, the only terms quadratic in $dz^a$ which are produced come only from $ds_{\cM_z}^2$ itself and thus must cancel amongst themselves. In principle this argument allows for $(\tbeta,\ta_c,\tb^c,\tB^c_{\ e})$ to depend on $(\phi,\sigma)$ as well but ultimately there is no such dependence. This is the main improvement of our expressions over those in \cite{deWit:1990na, deWit:1992wf}, in particular for symmetric spaces the matrix $\cD\ul{S}$ is explicitly independent of the coordinates $z^a$ of the base $\cM_z$, it depends only on $(\xi^A,\txi_A)$.

%%%%%%%%%%%%%%%%%%%%%%%%%%%%%%%%%%%%
\subsubsection{Cubic Prepotentials}
%%%%%%%%%%%%%%%%%%%%%%%%%%%%%%%%%%%%
For cubic prepotentials
\begin{equation}
	\cG = - D_{abc}\, \frac{Z^a Z^b Z^c}{Z^0}
\end{equation} 
we have 
\be
h(\xi^A, \txi_A)=\cI_4(\xi) \label{hI4}
\ee
where $\cI_4$ is the familiar quartic invariant defined in \eq{quarticdef}.  With this form of $h$ we note that $\underline{S}$ has the same form as $\UU$ in \eq{SymplAction} and \eq{QRSTdef} but its entries $(\tbeta,\ta,\tb,\tB)$ are now field dependent:
\bea
&& \tbeta= -\frac{1}{2} \Blp 3\txi_0 \xi^0 + \txi_a \xi^a \Brp\,, \quad\quad 
\tB_b^{\ a}=-\frac{1}{2}\Blp \frac{2}{3}\delta_b^a \txi_c \xi^c+ 2 \txi_b \xi^a -\frac{9}{8} \hD^{ace}D_{bcf} \xi^f \txi_e \Brp \label{betaBtilde}\\
&& \ta_a=-\frac{1}{2}\Blp 2 \xi^0 \txi_a+6 D_{abc}\xi^b \xi^c \Brp \,, \quad\quad 
\tb^a= -\frac{1}{2} \Blp 2 \txi_0 \xi^a - \frac{3}{32} \hD^{abc} \txi_b \txi_c\Brp \,.  \label{abtilde}
\eea
In a sense this is our main addition to the work of \cite{deWit:1990na, deWit:1992wf}, in that we provide the explicit form of the duality transformation on $\cM_z$ contained within the hidden Killing vectors on $\cM_h$. 

%%%%%%%%%%%%%%%%%%%%%%%%%%%%%%%%%%%%
\subsubsection{Quadratic Prepotentials}
%%%%%%%%%%%%%%%%%%%%%%%%%%%%%%%%%%%%
With base special K\"ahler manifold
\begin{equation}
	\cM_z=\frac{SU(1,n_h-1)}{U(1)\times SU(n_h-1)}
\end{equation} 
and quadratic prepotential
\begin{equation}
	\cG = Z^A \eta_{AB} Z^B, \qquad
	\eta = \frac{1}{2i} \diag(1, -1, \ldots, -1),
\end{equation} 
the resulting quaternionic K\"ahler manifold is the homogeneous space
\be
\cM_h=\frac{SU(2,n_h-1)}{SU(2)\times SU(n_h-1)\times U(1)}\,.
\ee
There is a unique quartic invariant and we find that
\be
h(\xi^A,\txi_A) =-\frac{1}{16} \cJ_4(\xi^A,\txi_A) 
\ee
where $\cJ_4$ is defined in \eq{J4def}. With this form we find that $\ul{S}$ has the form of \eq{UUcomponents} subject to \eq{quadSR} and \eq{quadQ} but with the non-trivial components having dependence on $(\xi^A,\txi_A)$. Explicitly we find
\bea
 \cR^{AB}&=& \half  \bslb \xi^A \xi^B +\frac{1}{8}(\eta^{-1})^{AB} (4 \xi \eta \xi - \txi \eta^{-1} \txi) -\frac{1}{4}(\eta^{-1} \txi)^A (\eta^{-1} \txi)^B  \bsrb  \label{Rxi}\\
 \cS_{AB}&=& \half  \bslb -\txi_A \txi_B +\frac{1}{2}\eta_{AB} (4 \xi \eta \xi - \txi \eta^{-1} \txi)+ 4(\eta \xi)_A (\eta\xi)_B  \bsrb \label{Sxi}
\eea
which satisfy \eq{quadSR} and also
\bea
 \cQ^{A}_{\ B} &=& \half  \bslb \xi^A \txi_B - (\eta^{-1} \txi)^A (\eta \xi)_B \bsrb\,,
 \eea
 which gives in components
 \bea
\cQ^A_{\ A}&=& 0\,,\quad\quad\cQ^{0}_{\ a} = \frac{1}{2} \blp \xi^0\txi_a +\delta^{0c} \delta_{ab} \txi_c \xi^b \brp\,,\quad\quad
\cQ^{a}_{\ b} =\frac{1}{2} \blp \xi^a \txi_b - \delta^{ac}\delta_{be}\txi_c \xi^e \brp  \label{Qxi}
\eea
and satisfies \eq{quadQ}.

This concludes our description of the Killing vectors on quaternionic K\"ahler manifolds. We next turn to the computation of the Killing prepotentials for these Killing vectors which will involve computing the compensators $W^x_\Lam$.

\subsection{Killing vector algebra}\label{sec:killing-algebra}

The non-vanishing commutators of the algebra are~\cite{deWit:1992wf}
\begin{equation}
	\label{eq:killing-algebra}
	\begin{gathered}
		\,[h_{\eps_0}, h_{\eps_+}] = 2 h_{\eps_+}, \qquad
		[h_{\eps_0}, h_\alpha] = h_\alpha, \qquad
		[h_\alpha, h_\alpha^t] = \mathbb C\, h_{\eps_+}, \qquad
		[h_{\mathbb U}, h_\alpha] = \mathbb{U}\, h_\alpha, \\
		[h_{\eps_0}, h_{\eps_-}] = -2\, h_{\eps_-}, \qquad
		[h_{\eps_0}, h_{\widehat\alpha}] = - h_{\widehat\alpha}, \qquad
		[h_{\eps_-}, h_\alpha] = -h_{\widehat\alpha}, \\
		[h_{\eps_+}, h_{\eps_-}] = - h_{\eps_0}, \qquad
		[h_{\eps_+}, h_{\widehat\alpha}] = h_\alpha, \qquad
		[h_{\mathbb U}, h_{\widehat\alpha}] = \mathbb{U}\, h_{\widehat\alpha}, \\
		[h_{\widehat\alpha}, h_{\widehat\alpha}^t] = \mathbb C\, h_{\eps_-}, \qquad
		[\widehat\alpha^t h_{\widehat\alpha}, \alpha^t h_\alpha] = \frac{1}{2}\, \widehat\alpha^t \mathbb C \alpha\, h_{\eps_0} + h_{\mathbb T_{\alpha, \hat\alpha}}
	\end{gathered}
\end{equation}
with
\begin{subequations}
\begin{gather}
	\mathbb T_{\alpha, \hat\alpha} = (\alpha^t \partial_\xi) (\hat\alpha^t \partial_\xi)\, \underline{S}
		= - \frac{1}{2}\, \mathbb C (\hat \alpha \alpha^t + \alpha \hat \alpha^t)
			+ \frac{1}{4}\, H''_{\alpha, \hat\alpha} \mathbb C, \\
	H''_{\alpha, \hat\alpha} = \partial_\xi (\partial_\xi h''_{\alpha, \hat\alpha})^t
		= (\alpha^t \partial_\xi) (\hat\alpha^t \partial_\xi) H, \\
	h''_{\alpha, \hat\alpha} = (\alpha^t \partial_\xi) (\hat\alpha^t \partial_\xi) h.
\end{gather}
\end{subequations}
There are two Heisenberg subalgebra, one generated by $\{ h_\alpha, h_{\eps_+} \}$, the
other by $\{ h_{\widehat\alpha}, h_{\eps_-} \}$.

\begin{figure}[h]
\begin{center}
	\label{fig:G2-root-diagram}
	\includegraphics[scale=1]{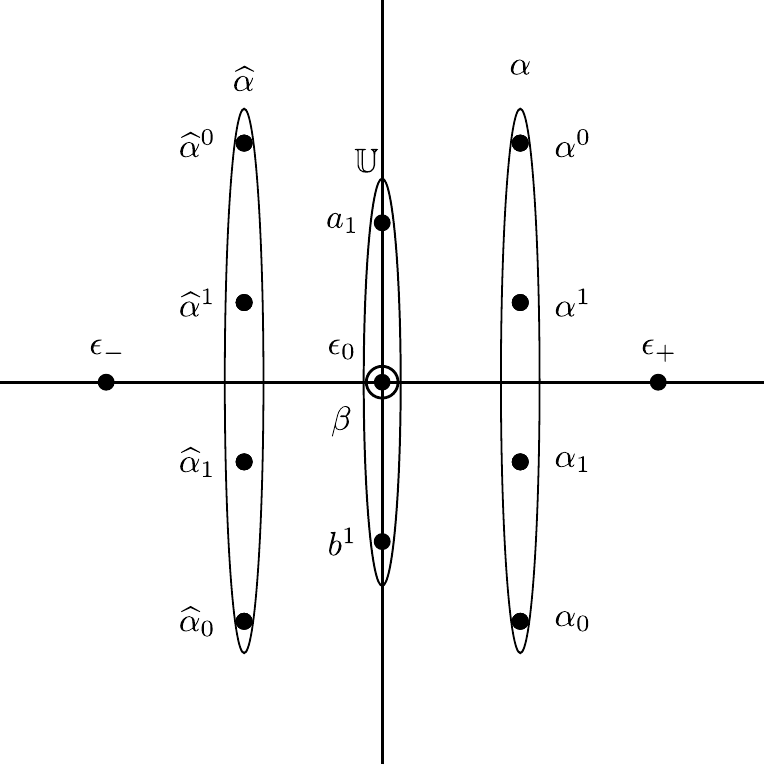}
	\caption{Root diagram for the $G_2$ Lie algebra.}
\end{center}
\end{figure}

As an example we give the $G_2$ root diagram in figure~\ref{fig:G2-root-diagram}. The hidden symmetries are on the left side.

%%%%%%%%%%%%%%%%%%%%%%%%%%%%%%%%%%%%
\section{The Killing prepotentials and compensators}
%%%%%%%%%%%%%%%%%%%%%%%%%%%%%%%%%%%%

For applications to $\cN=2$ gauged supergravity we need to compute the Killing prepotentials. As reviewed in section \ref{app:quaternionic} the action of a Killing vector on the spin connection may induce a local $SU(2)$ transformation:
\bea
\cL_\Lam(\om^x) = dW^x_\Lam + \eps^{xyz} W^y_\Lam \om^z
\eea
and $W^x_\Lam$ is then referred to as a {\it compensator}. The Killing prepotentials are then given by
\be
P^x_\Lam = k_\Lam \lrcorner \om^x - W^x_\Lam
\ee
and we find this to be an efficient route to computing the Killing prepotentials. We use the canonical expression for the spin connection using homogeneous coordinates on $\cM_z$ \cite{Ferrara:1989ik}\footnote{this expression is of course not invariant under local $SU(2)$ transformations and neither are our expressions for $P^x_\Lam$ or $W^x_\Lam$}:
\bea
\om^+&=& \sqrt{2}  e^{\frac{K_\Om}{2}+\phi} Z^T  \CC d\xi \\
\om^3&=&\frac{e^{2\phi}}{2} \blp d\sig + \half   \xi^T \CC d\xi \brp + \frac{1}{2}  e^{K_\Om}  \bslb\Gbar_{B}dZ^B - \Zbar^A dG_A +c.c.\bsrb
\eea
although one is of course free to choose another gauge. We have denoted the K\"ahler potential on $\cM_z$ by $K_{\Om}$.
%%%%%%%%%%%%%%%%%%%%%%%%%%%%%%%%%%%%
\subsection{The Compensators}
%%%%%%%%%%%%%%%%%%%%%%%%%%%%%%%%%%%%

%%%%%%%%%%%%%%%%%%%%%%%%%%%%%%%%%%%%
\subsubsection{Duality Symmetries}\label{sec:DualitySymm}
%%%%%%%%%%%%%%%%%%%%%%%%%%%%%%%%%%%%

We find that the spin connection is exactly invariant under all the duality symmetries except some components of $h_{\UU}$. For the quaternionic K\"ahler manifolds where $\cM_z$ has a cubic prepotential, we find 
 \be
\cL_{\UU}(\om^+) = -i a_c \Im z^c \, \om^+\,.
 \ee
 and so the only non-trivial compensator is
\be
W_{\UU}^3= a_c  \Im z^c\,.
\ee
For the $\cM_h$ where $\cM_z$ has a quadratic prepotential, we find 
\be
\cL_{\cQ^a_{\, 0}} \blp \om^+ \brp =-i \, \Im z^a\,  \cQ^a_{\ 0} \om^+ \,,\quad\quad
\cL_{\cR^{a0}} \blp \om^+ \brp =-i \,\Re z^a\,  \cR^{a0} \om^+
\ee
and thus the non-trivial compensators for the duality symmetries
\be
W^3_{\cQ^{a}_{\ 0}} = \Im z^a\,  \cQ^a_{\ 0}\,,\quad\quad 
W^3_{\cR^{a0}} = \Re z^a\,  \cR^{a0} \,.
\ee

%%%%%%%%%%%%%%%%%%%%%%%%%%%%%%%%%%%%
\subsubsection{Hidden Symmetries}
%%%%%%%%%%%%%%%%%%%%%%%%%%%%%%%%%%%%

For the hidden symmetries all components of the compensator are non-trivial. Nonetheless we can derive an expression which is equally valid for all prepotentials since the model dependence appears only through the compensator for the duality symmetry $h_{\UU}$.
\begin{equation}
	\begin{array}{rcl}
	W_{\eps_-}^+ &=& -i 2\sqrt{2} \, e^{\frac{K_\Om}{2} - \phi} \, Z^T  \CC \xi  \\
	W^3_{\eps_-}&=& - W^3_{\ul{S}} - e^{-2\phi}  \\
	W^+_{\hal}&=& - \CC \del_\xi W^+_{\eps_-} = i 2\sqrt{2} \, e^{\frac{K_\Om}{2}-\phi} \, \CC Z \\
	W^3_{\hal}&=& -2\CC \del_\xi W^3_{\eps_-}
	\end{array}\,.
\end{equation} 
The expression $W^x_{\ul{S}}$ is defined to mean $W^x_{\UU}$ with the parameters in $\UU$ promoted to the field dependent quantity $\ul{S}$ using \eq{betaBtilde}-\eq{abtilde} in the cubic case and \eq{Rxi}-\eq{Qxi} in the quadratic case. Similarly to the Killing vectors, $W_{\hal}$ is a $2n_h$-dimensional vector.

%%%%%%%%%%%%%%%%%%%%%%%%%%%%%%%%%%%%
\subsection{Killing Prepotentials}
%%%%%%%%%%%%%%%%%%%%%%%%%%%%%%%%%%%%

We find the Killing prepotentials by using
\be
P^x_\Lam = k_\Lam \lrcorner \om^x - W^x_\Lam\,.
\ee
Since we have already computed the compensators, it remains to just  compute $k_\Lam \lrcorner \om^x$ for the various Killing vectors. This contraction must be done in special co-ordinates, not homogeneous co-ordinates.

For the universal symmetries we have
\begin{equation}
	\begin{array}{ll}
	P^+_{\eps_+} = 0\,,\quad\quad\quad\quad& P^3_{\eps_+} = \half e^{2\phi} \,, \\
	P^+_{\eps_0} = \frac{1}{\sqrt{2}} e^{\frac{K_\Om}{2} + \phi} Z^T  \CC\xi \,,\quad\quad\quad\quad& P^3_{\eps_0} =\half e^{2\phi} \sig,   \\
	P^+_{\al} = - \sqrt{2}  e^{\frac{K_\Om}{2} + \phi} \CC Z  \,, \quad\quad\quad\quad& P^3_{\al} = - \frac{1}{2} e^{2\phi} \CC \xi
	\end{array}
\end{equation} 
For the model-dependent symmetries on the special K\"ahler base the prepotentials are
\begin{equation}
	\begin{array}{ll}
	P^+_{\UU} =  \sqrt{2} e^{\frac{K_{\Om}}{2} + \phi}Z^T \CC \UU \xi\,, \quad\quad 
	& P^3_{\UU} = \frac{1}{4} e^{2\phi} \xi^T \CC \UU \xi -   e^{K_\Om} Z^T \CC \UU \Zbar
	\end{array}
\end{equation}
For the hidden symmetries we find
\begin{equation}
	\begin{array}{l}
	P^+_{\eps_-}  =  \sqrt{2}  e^{\frac{K_\Om}{2} + \phi} \Bslb \sig Z^T  \CC \xi -  i 2 e^{-2\phi} \xi^T \CC Z- Z^T \CC (\del_\xi  W) \Bsrb \\ 
	P^3_{\eps_-} = \frac{1}{2} e^{-2\phi} +\frac{\sig^2}{2} e^{2\phi} -\frac{1}{4} e^{2\phi} \Bslb 2 W + \xi^T \CC (\del_\xi W) \Bsrb+ e^{K_\Om} \Zbar^T \CC\ul{S} Z
	\end{array}
\end{equation}
and
\begin{equation}
	\begin{array}{l}
	P^+_{\hal} =  -\sqrt{2} e^{\frac{K_\Om}{2} + \phi} (Z^T\CC\xi ) \CC \xi - 2 \CC (\del_\xi P^+_{\eps_-} ) \\
	P^3_{\hal}=- \CC\Bslb \sigma\,e^{2\phi} \xi+2 \del_\xi P^3_{\eps_-} \Bsrb
	\end{array}
\end{equation}

%%%%%%%%%%%%%%%%%%%%%%%%%%%%%%%%%%%%
\section{The Gauging}\label{sec:gaugings}
%%%%%%%%%%%%%%%%%%%%%%%%%%%%%%%%%%%%

Once the Killing vectors are classified a gauged supergravity theory is specified by a large set of gauging parameters which dictate how the various fields are charged. In this section we present the constraints on the embedding tensor for our Abelian gaugings. We denote the set of all Killing vectors of the hypermultiplets by
\begin{equation}
	k_\cA = \{ h_{\mathbb U}, h_\alpha, h_{\widehat\alpha}, h_{\epsilon_+}, h_{\epsilon_0}, h_{\epsilon_-} \}
\end{equation} 
and consider the most general gauging by introducing electric and magnetic parameters
\begin{equation}
	\label{eq:gauging-parameters}
	\Theta^\cA =
	\begin{pmatrix}
		\Theta^{\cA\Lambda} \\
		\Theta^\cA_\Lambda
	\end{pmatrix}
\end{equation} 
for each of these Killing vectors
\begin{equation}
	\Theta^\cA = \{ \mathbb U, \alpha, \widehat\alpha, \epsilon_+, \epsilon_0, \epsilon_- \} \label{gaugepars}
\end{equation} 
where we allow for a different symmetry $\UU_{\Lam}$ for each vector field. 
Each of the parameters is a symplectic vector whose components are of the same dimension than the corresponding Killing vectors (see appendix~\ref{app:gauging-constraints} for explicit lists). In particular all the parameters of the matrix $\mathbb U$ become symplectic vectors.

Contracting the Killing vectors with the parameters give the Killing vectors $\cK$
\begin{equation}
	\label{eq:def-k}
	\cK = \cK^u \frac{\del}{\del q^u}
		= \Theta^\cA k_\cA
		= h_{\mathbb U}
			+ \alpha^t \mathbb C h_\alpha
			+ \widehat\alpha^t \mathbb C h_{\widehat\alpha}
			+ \epsilon_+ h_{\epsilon_+}
			+ \epsilon_0 h_{\epsilon_0}
			+ \epsilon_- h_{\epsilon_-}
\end{equation} 
that couple to the electric and magnetic gauge fields~\footnote{Our notation is not very convenient for the vector $h_{\mathbb U}$: by the contraction we mean that the matrices $\mathbb U^\Lambda$ and $\mathbb U_\Lambda$ are used as the parameter for $h_{\mathbb U}$, i.e. we have $h_{\mathbb U^\Lambda}$ and $h_{\mathbb U_\Lambda}$.}. Splitting the electric and magnetic components give
\bea
	\label{eq:def-k-electric}
	k_\Lambda &=& k^u_\Lam \frac{\del}{\del q^u}
		= h_{\mathbb U_\Lambda}
			+ \alpha^t_\Lambda \mathbb C h_\alpha
			+ \widehat\alpha^t_\Lambda \mathbb C h_{\widehat\alpha}
			+ \epsilon_{+\Lambda} h_{\epsilon_+}
			+ \epsilon_{0\Lambda} h_{\epsilon_0}
			+ \epsilon_{-\Lambda} h_{\epsilon_-}, \\
	\label{eq:def-k-magnetic} 
	\widetilde k^\Lambda &=& \tk^{\Lam\,u} \frac{\del}{\del q^u}
		= h_{\mathbb U^\Lambda}
			+ \alpha^{t \Lambda} \mathbb C h_\alpha
			+ \widehat\alpha^{t \Lambda} \mathbb C h_{\widehat\alpha}
			+ \epsilon_+^\Lambda h_{\epsilon_+}
			+ \epsilon_0^\Lambda h_{\epsilon_0}
			+ \epsilon_-^\Lambda h_{\epsilon_-}
\eea
Electric and magnetic gaugings are distinguished only by the position of their $\Lambda$ index.

The number of parameters is
\begin{equation}
	\#(\text{params}) = 2 n_v \times \big[ (4 + x) n_h + 3 \big]
\end{equation} 
since for each of the $(2 n_v)$-dimensional symplectic vector component there is: $3$ parameters for $h_{\epsilon_0}$ and $h_{\epsilon_\pm}$, $2 n_h$ parameters for $\alpha$ and $\widehat \alpha$ and $x n_h$ parameters for $h_{\mathbb U}$ ($x$ being of order $1$ or $n_h$ depending on the model under consideration). All these parameters are not independent since consistency impose relations between them.

%%%%%%%%%%%%%%%%%%%%%%%%%%%%%%%%%%%%
\subsection{Constraints on the gauging parameters}\label{sec:gaugingconstraints}
%%%%%%%%%%%%%%%%%%%%%%%%%%%%%%%%%%%%

The gauging parameters are constrained by two conditions~\cite{deWit:2005ub, Samtleben:2008pe, deWit:2011gk}: closure of the Killing vector algebra
\begin{equation}
	\label{eq:gauging-algebra}
	[k_\cA, k_\cB] = f_{\cA\cB}^{\;\cC}\, k_\cC
\end{equation}
and locality. These two conditions are also necessary for satisfying the supersymmetric Ward identities~\cite{Samtleben:2008pe}.

Since only the hypermultiplet isometries are gauged, the Killing vectors $k_\cA$ form an abelian algebra~\cite{deWit:2011gk, Cassani:2012pj}. As a consequence the following commutators need to vanish
\begin{equation}
	[k_\Lambda, k_\Sigma] = [k_\Lambda, \widetilde k_\Sigma] = [\widetilde k_\Lambda, \widetilde k_\Sigma] = 0.
\end{equation} 

Upon inserting the explicit expression \eqref{eq:def-k-electric} of $k_\Lambda$ and using the algebra \eqref{eq:killing-algebra}, the first commutator leads to a set of quadratic constraints (see \eq{TTensorDef} for the definition of $\mathbb{T}$)
\begin{subequations}
\begin{align}
	0 &= \mathbb T(\alpha_\Lambda, \hat\alpha_\Sigma) - \mathbb T(\alpha_\Sigma, \hat\alpha_\Lambda), \\
	0 &= - (\mathbb U_\Lambda \alpha_\Sigma
			- \mathbb U_\Sigma \alpha_\Lambda)
		+ (\epsilon_{0\Lambda} \alpha_\Sigma
			- \epsilon_{0\Sigma} \alpha_\Lambda)
		+ (\epsilon_{+\Lambda} \widehat\alpha_\Sigma
			- \epsilon_{+\Sigma} \widehat\alpha_\Lambda), \\
	0 &= (\mathbb U_\Lambda \widehat\alpha_\Sigma
			- \mathbb U_\Sigma \widehat\alpha_\Lambda)
		+ (\epsilon_{-\Lambda} \alpha_\Sigma
			- \epsilon_{-\Sigma} \alpha_\Lambda)
		+ (\epsilon_{0\Lambda} \widehat\alpha_\Sigma
			- \epsilon_{0\Sigma} \widehat\alpha_\Lambda), \\
	0 &= \alpha^t_\Lambda \mathbb C \alpha_\Sigma
		+ 2 (\epsilon_{+\Sigma} \epsilon_{0\Lambda}
			- \epsilon_{+\Lambda} \epsilon_{0\Sigma}), \\
	0 &= (\widehat\alpha^t_\Lambda \mathbb C \alpha_\Sigma
			- \alpha^t_\Lambda \mathbb C \widehat\alpha_\Sigma)
		+ 2 (\epsilon_{+\Sigma} \epsilon_{-\Lambda}
			- \epsilon_{+\Lambda} \epsilon_{-\Sigma}), \\
	0 &= \widehat\alpha^t_\Lambda \mathbb C \widehat\alpha_\Sigma
		+ 2 (\epsilon_{0\Lambda} \epsilon_{-\Sigma}
			- \epsilon_{0\Sigma} \epsilon_{-\Lambda}).
\end{align}
\end{subequations}
These constraints involves product of electric parameters. The two other commutators lead to similar constraints for electric/magnetic and magnetic/magnetic products (see appendix \ref{app:gauging-constraints}).

The so-called locality constraints implies that the electric/magnetic duality exists and that we can rotate to a frame which is purely electric. Using the notation \eqref{eq:gauging-parameters} for the gauging parameters this condition reads
\begin{equation}
	\langle \varTheta^\alpha, \varTheta^\beta \rangle = 0.
\end{equation} 
The explicit list is given in appendix~\ref{app:gauging-constraints}. These constraints generalize the one given in~\cite{Cassani:2012pj}.

A consequence of the locality constraints is that the symplectic product of $\cK^u$ with $\cP^x$ always vanishes
\begin{equation}
	\label{eq:product-K-P}
	\langle \cK^u, \cP^x \rangle = 0.
\end{equation} 
The prepotential is linear in the gauging parameters and it can be written
\begin{equation}
	\cP^x = \varTheta^\cA P^x_\cA.
\end{equation} 
Inserting this expression and \eqref{eq:def-k} into the brackets we get
\begin{equation}
	\langle \cK^u, \cP^x \rangle = \cK^u_\cA \cP^x_\cB \langle \varTheta^\cA, \varTheta^\cB \rangle = 0.
\end{equation} 

%%%%%%%%%%%%%%%%%%%%%%%%%%%%%%%%%%%%%%
\section{Examples}
%%%%%%%%%%%%%%%%%%%%%%%%%%%%%%%%%%%%%%

In this section we work through two examples of gauged supergravity theories which arise from M-theory and which have $\cM_h=G_{2(2)}/SO(4)$, reproducing the $\cN=2$ AdS$_4$ vacuum and then look at black hole horizons.  It is well known that when an FI-gauged supergravity theory (i.e. with $n_h=0$ and $U(1)_R$ gauging) admits an $\cN=2$ AdS$_4$ vacuum it also admits a constant scalar flow to AdS$_2\times \HH_2/\tGam$, one can find a very general proof of this in \cite{Dall'Agata:2010gj}. With the addition of hypermultiplets, one can set them also constant and then the only additional constraints are $\langle \cK^u,\cQ \rangle=0$. Subject to this being solved, the hypermultiplets decouple and the constant scalar flow is also a solution of the theory with hypermultiplets. We demonstrate this in our two examples.

Our first example was obtained in \cite{Cassani:2012pj} corresponding to the invariant dimensional reduction of M-theory on $V_{5,2}$.  Our second example comes from \cite{Donos:2010ax} and corresponds to a consistent truncation of the dimensional reduction of maximal gauged supergravity on the Einstein three-manifold\footnote{$\Gamma$ is a discrete subgroup of $SL(2,\CC)$} $M_3\in \{ \HH_3/\Gamma ,T^3, S^3\}$. 
%%%%%%%%%%%%%%%%%%%%%%%%%%%%%%%%%%%%%%
\subsection{\texorpdfstring{$V_{5,2}$}{V5,2}}
%%%%%%%%%%%%%%%%%%%%%%%%%%%%%%%%%%%%%%
The invariant reduction of M-theory on seven-dimensional cosets was performed in \cite{Cassani:2012pj} where in addition the general reduction on $SU(3)$-structure manifolds was performed. All the resulting four dimensional gauged supergravity models found in that work fall into the class studied here, namely the hypermultiplet scalar manifold is a symmetric space which lies in the image of a c-map. Black hole solutions in many of these models were studied in \cite{Halmagyi:2013sla}, here we restrict ourselves to the example where $\cM_h=G_{2(2)}/SO(4)$ corresponding to the reduction on $V_{5,2}$. 

The following data specifies the four dimensional supergravity theory~\cite{Cassani:2012pj}:
\bea
&&n_v=1\,,\quad\quad \cM_v=\frac{SU(1,1)}{U(1)}\,,\quad\quad \cF=-\frac{(X^1)^3}{X^0}\,,\quad\quad X^\Lam=\bpm1 \\ \tau\epm\,, \\
&&n_h=2\,,\quad\quad \cM_h=\frac{G_{2(2)}}{SO(4)}\,,\quad\quad \cM_z=\frac{SU(1,1)}{U(1)}\quad\quad \cG=-\frac{(Z^1)^3}{Z^0}\,,\quad\quad Z^\Lam =\bpm 1 \\ z \epm\,.
\eea
The nonvanishing electric gaugings are given by
\be
b^1_\Lam=\frac{4}{\sqrt{3}} \delta_{\Lam 0}\,,\quad a_{1,\Lam}=-\frac{4}{\sqrt{3}} \delta_{\Lam 0}\,,\quad \eps_{+\Lam}= -e_0 \delta_{\Lam0}\,.
\ee
The non-vanishing magnetic gauging is given by
\bea
&&\eps_+^\Lam=-2 \delta^{\Lam1}
\eea
The constant $e_0$ has its origin in the M-theory three-form with legs in the external four dimensional spacetime which has been dualized to a constant \cite{Cassani:2012pj}.

We note that the gaugings which specify this model were incorrectly reported in \cite{Cassani:2012pj} to have vanishing compensator $W^x_\Lam$. This of course is incompatible with the existence of a supersymmetric AdS$_4$ vacuum. The resolution is that as found in section \ref{sec:DualitySymm} the Killing vectors $k_\UU$ with $a_i\neq0$ have non-trivial compensators and we now see this is nontrivially gauged. In fact this is the only gauging with a non-trivial compensator in this reduction.

%%%%%%%%%%%%%%%%%%%%%%%%%%%%%%%%%%%%%%
\subsubsection{AdS$_4$ Vacua}
%%%%%%%%%%%%%%%%%%%%%%%%%%%%%%%%%%%%%%

The Killing prepotentials $P^\pm_\Lam$ are set to vanish by the condition
\be
\xi^A=\txi_A=0\,.
\ee
Then from $\langle \cK^{\Adot},\Im \cV \rangle=0$ (in the direction of $\cM_z$) we get
\be
\cK^{\Adot}=0\quad \Rightarrow\quad z^1=i\sqrt{3}\,.
\ee
and from  $\langle \cK^a,\Im \cV \rangle=0$ (in the direction of the axion $a$) we get
\be
e^{\phi}=\sqrt{\frac{6}{e_0}}
\ee
while the axion is unfixed. As a result we have the Killing prepotentials
\be
P_\Lam^3=( 1\,,0)\,,\quad\quad
\tP^{3,\Lam}=(  0,-6/e_0)\,.
\ee
The vector multiplet scalars are then given by
\be
x=0\,,\quad\quad y=\sqrt{\frac{e_0}{6}}
\ee
and the AdS$_4$ radius is given by
\be
R^2_{{\rm AdS}_4}=\frac{12\sqrt{6}}{e_0^{3/2}}\,.
\ee

%%%%%%%%%%%%%%%%%%%%%%%%%%%%%%%%%%%%%%
\subsubsection{AdS$_2\times \Sig_g$ Vacua}
%%%%%%%%%%%%%%%%%%%%%%%%%%%%%%%%%%%%%%

There is a related AdS$_2\times \HH^2/\tGam$ vacuum at the same point on the scalar moduli spaces $\cM_v\times \cM_h$. The charges are
\be
\cQ=(\frac{1}{4},0,0,\frac{e_0}{8})
\ee
and the radii are 
\be
R_1= \frac{e_0^{3/4}}{8 ( 2^{1/4}3^{3/4})}\,,\quad\quad R_2= \frac{e_0^{3/4}}{4 (2^{1/4}3^{3/4})}\,.
\ee

%%%%%%%%%%%%%%%%%%%%%%%%%%%%%%%%%%%%%%
\subsection{\texorpdfstring{$SO(5)$}{SO5} Gauged Supergravity on \texorpdfstring{$M_3$}{M3}}
%%%%%%%%%%%%%%%%%%%%%%%%%%%%%%%%%%%%%%
The maximal gauged supergravity in seven dimensions \cite{Pernici:1984xx} has been dimensionally reduced on three-dimensional constant curvature Einstein manifolds and consistently truncated to a four dimensional gauged supergravity theory in \cite{Donos:2010ax}. The resulting theory is given by the following data:
\bea
&& n_v=1\,,\quad \cM_v=\frac{SU(1,1)}{U(1)}\,,\quad \cF=-4\frac{(X^1)^3}{X^0}\,,\quad\quad X^\Lam=\bpm1 \\ \tau\epm\,, \\
&&n_h=2\,,\quad\quad \cM_h=\frac{G_{2(2)}}{SO(4)}\,,\quad\quad \cM_z=\frac{SU(1,1)}{U(1)}\quad\quad \cG=-\frac{(Z^1)^3}{Z^0}\,,\quad\quad Z^\Lam =\bpm 1 \\ z \epm\,.
\eea
We have computed the gaugings in our terminology by careful comparison with \cite{Donos:2010ax}. This requires a non-trivial co-ordinate change which is detailed in appendix \ref{app:G2}.

To specify the gaugings we need only to give the components of the embedding tensor in \eq{gaugepars}.  We find that $k_1=0$ and the non-vanishing electric components are in $k_0$
\be
\al^0_{\ ,0}=\frac{1}{2}\,,\quad\quad \hal_{0,0}= 3^{3/4}\,,\quad\quad \al_{1,0} =\frac{3^{3/4 }\ell}{4}\,.
\ee
Likewise we find that $\tk^0=0$ and the non-vanishing magnetic components are in $\tk^1$
\be
\al_{1,}^{\ 1}=-\frac{1}{2\sqrt{3}}\,.
\ee
The integer $\ell=\{-1,0,1\}$ corresponds to the reduction on $M_3=\{\HH_3/\Gamma,T^3,S^3\}$ respectively. The gauging from $\hal_{0,0}$ provides the non-trivial compensator required to have a supersymmetric AdS$_4$ vacuum.

This yields the  magnetic Killing prepotentials
\be
\tP^{x,0}=0\,,\quad\quad \tP^{1,1}= \frac{3^{1/4}}{2} e^{\phi+3\vphi} \chi \,,\quad\quad 
\tP^{2,1} = \frac{3^{1/4}}{2} e^{\phi+\vphi} \,,\quad\quad
\tP^{3,1} =  \frac{3^{1/4}}{2} e^{2\phi} \xi^1
\ee
and the electric Killing prepotentials
\bea
P^1_0&=& \frac{1}{3^{3/4}4 } \Bslb -9 e^{4\vphi} \chi \ell +2\chi (e^{4\vphi } \chi^2-3) \non \\
&&+ 3^{3/2} \Blp 6 \xi^0 (\xi^1-\chi \xi^0) + e^{4\vphi} \blp -2\sig + \xi^0 (\txi_0+2\chi^3 \xi^0)+ \txi_1 \xi^1 - 6\chi^2 \xi^0 \xi^1 + 6 \chi (\xi^1)^2 \brp \Brp \Bsrb \non \\
P^2_0&=& \frac{1}{3^{3/4}4} \Bslb -9 e^{\phi+\vphi} \ell +2 e^{-\phi-3\vphi} \Blp e^{2\phi} (3e^{4\vphi} \chi^2-1) +3^{3/2} \blp  e^{2\phi} (3 e^{4\vphi} (-\chi \xi^0 +\xi^1)^2)-(\xi^0)^2 -e^{6\vphi}\brp \Brp  \Bsrb \non \\
&& \\%
P^3_0&=& \frac{1}{3^{3/4}4} \Bslb 18 \sqrt{3} e^{2\vphi}(\chi \xi^0-\xi^1) +e^{2\phi} \blp \txi_0(2+3^{3/2}(\xi^0)^2)    -9 \ell \xi^1 + 3^{3/2} (\txi_1 \xi^0 \xi^1 +2(\xi^1)^3-2\sig \xi^0) \brp \Bsrb\non \\
&& \non \\
P^x_1&=& 0\,. \non 
\eea

%%%%%%%%%%%%%%%%%%%%%%%%%%%%%%%%%%%%%%
\subsubsection{AdS$_4$ Vacua}
%%%%%%%%%%%%%%%%%%%%%%%%%%%%%%%%%%%%%%
The supersymmetric AdS$_4$ vacuum is at
\be
 \xi^A=\txi_A=\chi=a=\phi=0\,,\quad\quad e^{\vphi}= \frac{1}{3^{1/4}}\,,\quad\quad \tau^1 = \frac{i}{2\sqrt{2}} 
\ee
and in particular requires $\ell=-1$, corresponding to a reduction on $\HH_3/\Gamma$. The AdS$_4$ radius is
\be
R_{AdS_4}=\frac{1}{\sqrt{2}}\,.
\ee
Evaluated at this vacuum the Killing prepotentials become
\be
P^1_\Lam=P^3_\Lam=\tP^{1,\Lam}=\tP^{3,\Lam}=0\,,\quad\quad P^2_0=-\frac{1}{4}\,,\quad\quad \tP^{2,2}=\frac{1}{2}\label{KillPreAdS4}
\ee

%%%%%%%%%%%%%%%%%%%%%%%%%%%%%%%%%%%%%%
\subsubsection{AdS$_2\times \Sig_g$ Vacua}
%%%%%%%%%%%%%%%%%%%%%%%%%%%%%%%%%%%%%%

The AdS$_2\times \Sig_g$ vacuum for is located at the same point on the scalar manifold. The charges are given by
\be
p^0=-1\,,\quad p^1=0\,,\quad q_0=0\,,\quad q_1=-\frac{3}{2}
\ee
The radii are given by
\be
R_1= \frac{1}{2^{3/4}}\,,\quad\quad R_2=\frac{1}{2^{1/4}}\,.
\ee
When lifted to M-theory this is a solution of the form 
\be
{\rm AdS}_2\times \HH^2/\tGam \times( \HH^3/\Gam \times_w S^4)
\ee
where the $S^4$ is fibered non-trivially over $\HH^3$.
It arises as the IR of a domain wall AdS$_4\ra$ AdS$_{2}\times \HH^2$ where the scalar fields take constant values along the whole flow. 
%%%%%%%%%%%%%%%%%%%%%%%%%%%%%%%%%%%%%%
\section{Conclusions}
%%%%%%%%%%%%%%%%%%%%%%%%%%%%%%%%%%%%%%

We have analyzed the symmetry structure of symmetric special quaternionic K\"ahler manifolds with a view towards studying general gaugings of $\cN=2$ supergravity. In particular we have computed the Killing prepotentials and compensators for all symmetries of such manifolds. We have shown in certain examples how this fits with existing theories in the literature derived from M-theory. 

The overarching goal of this study is a comprehensive understanding of BPS vacua in $\cN=2$ gauged supergravity, in particular black hole solutions. A particular goal, yet to be realized is to generalize the solution of black hole horizons in \cite{Halmagyi:2013qoa} to include hypermultiplets. This requires a deeper analysis of \eq{KVzero} and \eq{AdS2Sigeq5} as well as the constraints on the embedding tensor in section~\ref{sec:gaugings}. 

An interesting related computation was performed in \cite{Cassani:2009na} regarding the analysis of $\cN=1$ AdS$_4$ vacua in the same theories we have studied in this work. The key difference is that for $\cN=1$ AdS$_4$ vacua one only gauges the Heisenberg shift symmetries and these have vanishing compensators. Nonetheless with this simplification the authors of \cite{Cassani:2009na} could derive very general classes of AdS$_4$ vacua in theory coupled to hypermultiplets whose scalar manifold lies in the image of a c-map. 

A more immediate a modest goal is to complete the analysis of black hole horizons of \cite{Halmagyi:2013sla} by expressing the scalar fields and radii in terms of the charges. One lesson from the study of FI-gauged supergravity in \cite{Halmagyi:2013qoa} was that while this inversion can be a formidable task in any given example, it is advantageous to maintain the symplectic covariance by studying general classes of theories simultaneously. The models studied in \cite{Halmagyi:2013sla} have a hypermultiplet scalar manifold $\cM_h$ whose base special K\"ahler manifold $\cM_z$ has a quadratic prepotential. This can be studied using the techniques from this work and should result in complete solution for the black hole horizons for all the models of \cite{Cassani:2012pj}. This should involved carefully considering the embedding of the Abelian gauge group into the symplectic group or equivalently solving the constraints in section \ref{sec:gaugings}. A simple model of AdS$_4$ vacua was solved in \cite{Halmagyi:2011xh} where very particular patterns were observed regarding the dependence if if the solution space on the gauge group and its embedding.

Another interesting direction is to find the analytic black hole solutions for models with hypermultiplets much like the analytic solutions in FI-gauged supergravity \cite{Cacciatori:2009iz, Gnecchi:2013mta, Halmagyi:2014qza}. The key step in finding the most general dyonic static black hole these FI-gauged supergravity theories was to posit the ansatz whereby a particular metric function was, much like the Demianski-Plebanski solution \cite{Plebanski:1976gy}, a quartic polynomial in the radius. This ansatz may help in generalizing such analytic solutions to hypermultiplet theories, it seems like a difficulty problem but any progress would be an interesting development.

One last issue is that the computations in this paper can most likely be generalized to include all homogeneous quaternionic K\"ahler manifolds, not just the symmetric ones. For these manifolds, the hidden Killing vectors are significantly more complicated but given that they have been explicitly computed in \cite{deWit:1990na} one imagines it to be possible to compute the associated Killing prepotentials, we leave this for future investigations.

\vskip 1cm
%%%%%%%%%%%%%%%%%%%%%%%%%%%%%%%%%%%%%%%%%%%%%
\noindent {\bf Acknowledgments} We would like to thank Bernard de Wit for useful conversations.
This work was conducted within the framework
of the ILP LABEX (ANR-10-LABX-63) supported by French state funds managed by the
ANR within the Investissements d'Avenir programme under reference ANR-11-IDEX-0004-02

%%%%%%%%%%%%%%%%%%%%%%%%%%%%%%%%%%%%%%
\begin{appendix}
%%%%%%%%%%%%%%%%%%%%%%%%%%%%%%%%%%%%%%

%%%%%%%%%%%%%%%%%%%%%%%%%%%%%%%%%%%%
\section{Special K\"ahler Geometry}\label{sec:special-geometry}
%%%%%%%%%%%%%%%%%%%%%%%%%%%%%%%%%%%%

We start with a brief summary of special K\"ahler geometry. The key ingredients are a K\"ahler manifold $\cM_v$ equipped with an $Sp(2n_v+2,\RR)$ bundle over it with sections
\be
X=\bpm X^\Lam \\ F_\Lam \epm\,,\quad\quad \Lam=0,\ldots ,n_v\,.
\ee
We will be primarily concerned in this paper with the so-called {\it very} special K\"ahler manifolds, which means there is a cubic prepotential
\be
F=-d_{ijk}\frac{ X^i X^j X^k}{X^0}\,,\quad\quad i=1,\ldots ,n_v\,.
\ee
The canonical complex coordinates $\tau^i = x^i + i\, y^i$ on $\cM_v$ are called {\it special coordinates} 
\be
X^\Lam = \bpm 1 \\ \tau^i\epm  \quad \Ra \quad F_\Lam =\bpm d_\tau \\ -3 d_{\tau,i} \epm\,.
\ee
The metric can be obtained from a K\"ahler potential $K$
\bea
e^{-K}&=& -i X^T \Om \Xbar=8d_y \\
g_{i\jbar}&=& \del_i \del_{\jbar} K
\eea
where $\Om$ is the $(2n_v+2)\times (2n_v+2)$ dimensional matrix $
\Om=\bpm 0 & 1\!\! 1 \\ -1\!\!1 &  0\epm$\,. 

We next introduce the operators which appear in the gauge field Lagrangian
\be
\cN_{\Lam\Sig}=\cR_{\Lam \Sig}+ i \cI_{\Lam \Sig}
\ee
There is a very useful projection operator 
\bea
\cM=\bpm \cI^{-1} \cR & -\cI^{-1} \\ \cI + \cR \cI^{-1} \cR& -\cR \cI^{-1} \epm
\eea
which satisfies
\be
\cM \cV= -i \cV\,,\quad\quad \cM U_i = i U_i
\ee 
where $\cV=e^{K/2} X$ and 
\be
U_i = D_i \cV = \del_i \cV + \half \del_i K \cV
\ee

The Riemann tensor on $\cM_v$ is given by
\bea
R_{\ jk}^{i\ \ l}&=& \delta^i_j \delta^l_k+\delta^i_k \delta^l_j - \frac{9}{16} \hd^{ilm}d_{mjk}
\eea
where
\bea
\hd^{ijk}&=& \frac{g^{il} g^{jm} g^{kn}d_{lmn}}{d_y^2} 
\eea
When $\cM_v$ is in addition a homogeneous space, the tensor $\hd^{ijk}$ has constant entries and satisfies certain useful identities
\bea
\hd^{ijk} d_{jl(m}d_{np)k}&=&\frac{16}{27} \Bslb \delta^{i}_{l} d_{mnp} + 3 \delta^{i}_{(m}d_{np)l} \bsrb \\
\hd^{ijk} d_{j(lm}d_{np)k}&=&\frac{64}{27}  \delta^{i}_{(m}d_{npl)}\,.  \\
\eea

The quartic invariant is defined using both $d_{ijk}$ and $\hd^{ijk}$:
\be
I_4(\cQ)=-(p^0q_0+p^iq_i)^2 -4q_0 d_{ijk} p^ip^jp^k +\frac{1}{16}p^0 \hd^{ijk}q_iq_jq_k + \frac{9}{16} d_{ijk} \hd^{ilm} p^j p^kq_l q_m\,.
\ee
From this we obtain a symmetric four index tensor
\be
I_4(\cQ)=\frac{1}{4!}t^{MNRS}Q_MQ_NQ_RQ_S
\ee
which is then used to define the derivative of $I_4$:
\be
I'_4(\cQ)_M= \frac{1}{3!} \Om_{MN}t^{NRST}Q_RQ_SQ_T\,.
\ee
We note that $I'_4$ can be used to relate the real and imaginary parts of the symplectic section $\cV$
\be
\Re \cV= -\frac{I'_4(\Im \cV)}{2\sqrt{I_4(\Im \cV)}}\,.\label{ReVImV}
\ee

We will often employ the shorthand notation
\bea
d_\tau=d_{ijk} \tau^i \tau^j \tau^k\,,\qquad d_{\tau,i }=d_{ijk} \tau^j \tau^k\,,\qquad d_{\tau,ij }=d_{ijk}\tau^k\,.
\eea

%%%%%%%%%%%%%%%%%%%%%%%%%%%%%%%%%%%%
\subsection{Quadratic Prepotential}\label{app:quadraticprep}
%%%%%%%%%%%%%%%%%%%%%%%%%%%%%%%%%%%%

The general quadratic prepotential is 
\be
\cF=X^\Lam \eta_{\Lam \Sig} X^\Sig\,.
\ee
Using an orthogonal matrix we can diagonalize $\eta$ then with a complex rescaling of $X^\Lam$ we can set 
\be
\eta=\frac{1}{2i}\diag\{1,-1,\ldots ,-1\}\,.
\ee 
We then choose special coordinates:
\bea
X^{\Lam}=\bpm 1 \\ \tau^i \epm\,,\quad\quad F_\Lam =2\eta_{\Lam\Sig} X^\Sig= i\bpm -1\\ \tau^i \epm
\eea
giving
\bea
e^{-K}&=& 2(1-|\tau^i|^2)\,,\quad\quad
g_{i\jbar}= \frac{\delta_{i\jbar}}{1-|\vec{\tau}|^2}
\eea
which is the maximally symmetric metric on
\be
\cM_z=\frac{SU(1,n_v)}{U(1)\times SU(n_v)}\,.
\ee
Taking the variation of $F_\Lam =2\eta_{\Lam\Sig} X^\Sig$ we get
\bea
2 \eta_{\Lam\Sig} (\cQ^\Sig_{\ \Delta} X^\Delta + \cR^{\Sig \Delta} F_\Delta) = \cS_{\Lam\Sig} X^\Sig - (\cQ^T)_\Lam^{\ \Sig} F_\Sig \\
\Rightarrow\quad 2 \eta_{\Lam\Sig} (\cQ^\Sig_{\ \Delta} X^\Delta +2 \cR^{\Sig \Delta} \eta_{\Delta \Upsilon} X^{\Upsilon}) = \cS_{\Lam\Sig} X^\Sig -2 (\cQ^T)_\Lam^{\ \Sig} \eta_{\Sig\Delta}X^\Delta
\eea
which gives
\bea
\eta_{\Sig(\Lam} \cQ^\Sig_{\ \Delta)} &=& 0 \label{Qconstraint}\\
\cS_{\Lam \Sig}&=&4\eta_{\Lam\Upsilon} \cR^{\Upsilon \Delta} \eta_{\Delta \Sig}\,.
\eea
Note that \eq{Qconstraint} gives
\be
\cQ^{\Lam}_{\ \Lam}=0\,,\quad\quad \cQ^0_{\ i}=\cQ^i_{\ 0}\,,\quad\quad \cQ^i_{\ j} = -\cQ^{j}_{\ i}
\ee
($\Lam$ indices are not summed).

The special coordinates $\tau^i$ transform as
\bea
\delta \tau^i
&=&\cA^{i}_{\ 0}
-  \tau^i\cA^{0}_{\ 0}
+\cA^{i}_{\ j} \tau^j 
- \tau^i \tau^j \cA^{0}_{\ j}
\eea
where
\be
\cA= \cQ+2 \cR \eta\,.
\ee
From this we see that we should remove $\Tr\, \cA$ or $\cA^{0}_{\ 0}$ since their action on $\tau^i$ is redundant, this is tantamount to removing  $\Tr\, \cR$ or $\cR^{0}_{\ 0}$. This leaves the components
\be
\cR: \half(n_v+1)(n_v+2)-1\,,\quad\quad \cQ: \half n_v(n_v-1)+n_v
\ee
giving $n_v^2+2n_v$ which agrees with the number of Killing vectors on $SU(1,n_v)/\bslb U(1)\times SU(n_v)\bsrb$ thus demonstrating that all Killing vectors come from the symplectic action \eq{SymplAction}.

The Lie derivative of the K\"ahler potential gives
\bea
e^K \cL_{\UU}\Blp e^{-K}\Brp &=& - \Bslb \tau^i \blp \ol{\cA}^i_{\ 0} - \taubar^i \taubar^j \ol{\cA}^0_{\ j}\brp - |\tau^i|^2 \ol{\cA}^0_{\ 0} + \tau^i \taubar^j \ol{\cA}^i_{\ j} +c.c. \Bsrb  e^{K} \non \\
&=& -2 \Bslb x^i {\cQ}^i_{\ 0} 
- 2i y^i  (\cR\eta)^i_{\ 0}  \Bsrb\,,
\eea
so the K\"ahler potential tranforms as 
\be
\cL_{\UU}(K)=f_\UU(\tau^i)+\fbar_\UU(\taubar^i)\,,\quad\quad f_\UU(\tau^i)=2\tau^i \ol{\cA}^i_{\ 0} .
\ee

%%%%%%%%%%%%%%%%%%%%%%%%%%%%%%%%%%%%
\section{Quaternionic K\"ahler Geometry} \label{app:quaternionic}
%%%%%%%%%%%%%%%%%%%%%%%%%%%%%%%%%%%%
Here we collect some facts about quaternionic K\"ahler geometry. The triplet of curvature two-forms $\Om^x$ are given by
\bea
\Om^x&=& D \om^x= d\om^x + \half \eps^{xyz} \om^y\w  \om^z
\eea
where $\om^x$ is the $SU(2)$-valued spin connection. For each Killing vector $k_\Lam$ one can construct the moment maps, or Killing prepotentials $P^x_\Lam$:
\bea
-k_\Lam \lrcorner \Om^x &=& D P^x_\Lam\,.
\eea

The curvature forms need not be precisely invariant under the action of $k_\Lam$ but may transform by a compensating local $SU(2)$ transformation
\be
\cL_k \Om^x= \eps^{xyz} \Om^y  W^z_\Lam \,,\quad\quad \cL_k \om^x = D W^x_\Lam\,.
\ee
Following \cite{D'Auria:1990fj} page 719, one can show that the Killing prepotentials are given by
\be
P^x_\Lam = k_\Lam \lrcorner \om^x  - W^x_\Lam
\ee
and in addition the compensator $W^x_\Lam$ satisfies
\be
\cL_\Lam W^x_\Sig -\cL_\Sig W^x_\Lam + \eps^{xyz} W^y_{\Lam} W^z_{\Sig} = f^\Delta_{\Lam \Sig} W^x_\Delta \,.
\ee

%%%%%%%%%%%%%%%%%%%%%%%%%%%%%%%%%%%%
\subsection{Special Quaternionic K\"ahler Geometry}  \label{app:special-quaternionic}
%%%%%%%%%%%%%%%%%%%%%%%%%%%%%%%%%%%%

In this work we are primarily concerned with quaternionic K\"ahler manifolds $\cM_h$ (of real dimension $4n_h$) which lie in the image of the c-map. Amongst other things, this means that $\cM_h$ has a base $(2n_h-2)$-dimensional base manifold $M_z$ which is special K\"ahler. For such manifolds the metric takes the form 
\bea
h_{uv} dq^u dq^v = d\phi^2 + g_{a\bbar} dz^a d\zbar^{\bbar} +\frac{1}{4} e^{4\phi }\blp d\sig + \half \xi^T \CC d\xi\brp^2 -\frac{1}{4}e^{2\phi} d\xi^T \CC \MM d\xi \label{quatmet}
\eea
where $a=1,\ldots , n_h-1$ and
\be
\CC=\bpm 0 & 1\!\!1 \\ -1\!\!1 & 0\epm
\ee
and $\MM$ is the equivalent of $\cM$ but for $\cM_z$
\bea
\MM=\bpm \cI^{-1} \cR & -\cI^{-1} \\ \cI + \cR \cI^{-1} \cR& -\cR \cI^{-1} \epm
\eea

On the base special K\"ahler manifold we denote the sections by
\bea
\cZ= \bpm Z^A \\ G_A\epm\,, \quad\quad A=0\,,\ldots ,n_h-1\,.
\eea
Will will generically assume there is a prepotential $\cG$ which thus satisfies $G_A=\del_A \cG$, special co-ordintaes on $\cM_z$ are given by
\be
Z^A= \bpm1 \\ z^a \epm\,. 
\ee
The canonical expression for the spin connection \cite{Ferrara:1989ik} uses homogeneous coordinates on $\cM_z$:
\bea
\om^+&=& \sqrt{2}  e^{\frac{K_\Om}{2}+\phi} Z^T  \CC d\xi\,, \label{spinpm}\\
\om^3&=&\frac{1}{2} e^{2\phi} \blp d\sig + \half   \xi^T \CC d\xi \brp + \frac{1}{2}  e^{K_\Om}  \bslb\Gbar_{B}dZ^B - \Zbar^A dG_A +c.c.\bsrb \label{spin3}
\eea
where we have denoted the K\"ahler potential on $\cM_z$ by $K_\Om$.

%%%%%%%%%%%%%%%%%%%%%%%%%%%%%%%%%%%%%%
\subsection{Hidden symmetries: field variations} \label{app:hidden-killing}
%%%%%%%%%%%%%%%%%%%%%%%%%%%%%%%%%%%%%%

Following \cite{deWit:1990na, deWit:1992wf} we denote the parameters for these symmetries as $(\eps_-,\hal^A,\hal_A)$ and variations associated to the Killing vectors \eqref{eq:hidden-killing} are
\bea
\delta \rho &=& 2\rho \bslb \sig \eps_- + \half \hal^T \CC \xi \bsrb \\
\delta \sig &=&\sig \bslb \sig \eps_- + \half \hal^T \CC \xi\bsrb-\rho^2 \eps_- -\cD W  \\
\delta \xi &=&   \xi\bslb \sig \eps_-+ \half \hal^T \CC \xi \bsrb + \sig \hal - \del_\xi \cD W  \\
\delta Z &=&\cD \underline{S}\, Z
\eea
with $\rho = e^{-2 \phi}$
\bea
\hal=\bpm \hal^A \\ \hal_A  \epm\,,\quad\quad\cD&=&\eps_- - \hal^T \CC \del_\xi \,,\quad\quad\quad
W= \frac{1}{4} h(\xi^A,\txi_A) -\frac{1}{2} \rho \xi^T \CC \MM \xi
\eea
and $\underline{S}$ is the symplectic matrix:
\bea
\underline{S}&=&\half  \Blp \xi \xi^T +\half   H\Brp \CC\,,\quad\quad
H= \bpm \del^I \del^J  h(\xi^A,\txi_A)&- \del^I \del_J h(\xi^A, \txi_A) \\ - \del_I \del^J h(\xi^A,\txi_A) & \del_I \del_J h(\xi^A, \txi_A) \epm = \del_\xi \, (\del_\xi h)^T.
\eea

Our expression for $\delta \sig$ differs from that found in \cite{deWit:1990na, deWit:1992wf} by a component in the final term $\cD W$. We have not been able to check that our expression for $\delta Z $ precisely agrees with the expressions there.

%%%%%%%%%%%%%%%%%%%%%%%%%%%%%%%%%%%%%%
\subsection{Computing the Compensators} \label{app:compensators}
%%%%%%%%%%%%%%%%%%%%%%%%%%%%%%%%%%%%%%

We now provide some details about how we computed the compensators $W^x_\Lam$ for the duality symmetries as well as the hidden symmetries. We do this by computing the Lie derivative of the spin connection then using 
\bea
\cL_\Lam (\om^\pm)&=& dW^\pm_\Lam \mp i \om^\pm W^3_\Lam \pm  i \om^3 W^\pm_\Lam  \label{LieSpinpm}\\ 
\cL_\Lam (\om^3)&=& dW^3_\Lam +\Im (\om^- W^+_\Lam) \,.\label{LieSpin3}
\eea
The key point is that we must use special coordinates on $\cM_z$ in the expressions \eq{spinpm} and \eq{spin3}. Some of these calculations are lengthy but in principle they are all fairly straightforward.

%%%%%%%%%%%%%%%%%%%%%%%%%%%%%%%%%%%%%%
\subsubsection{Duality Symmetries}
%%%%%%%%%%%%%%%%%%%%%%%%%%%%%%%%%%%%%%

Under the Cartan transformation $\beta$ we have
\be
\cL_\beta (e^{K_\Om/2}) = \beta e^{K_\Om/2} \label{cLOm2}
\ee
as well as
\bea
\cL_\beta \Blp Z^T \CC  d\xi\Brp &=&-\beta d\txi_0- \frac{\beta}{3!} D_z d\xi^0 -\frac{\beta}{3} z^a d\txi_a + \frac{\beta}{6} D_{z,a} d\xi^a  +2\beta \frac{1}{3!} D_z d\xi^0-\frac{2\beta }{3} z^a d\txi_a- \frac{2\beta}{3} D_{z,a} d\xi^a \non \\
&=& -\beta Z^T \CC  d\xi\,.\label{cLZCxi}
\eea
In total \eq{cLOm2} and \eq{cLZCxi} give
\be
\cL_{\beta}(\om^+)=0
\ee
and this demonstrates the need to compute in special co-ordinates. Similarly one finds 
\be
\cL_\beta(\om^3)=0
\ee

Under the $a_c$-symmetries the special coordinates transform as
\bea
\delta z^a& =& - \half R^{a\ \ \, e}_{\ bc}z^b z^c  a_e = -z^a (a_e z^e) + \frac{9}{32} a_e \hD^{aef}D_{z,f}
\eea
and we find 
\bea
\cL_a(e^{K/2}) &=& a_c \Re z^c\, e^{K/2}\,,\quad\quad \cL_{a} (Z^T \CC d\xi ) = -a_c z^c Z^T \CC d\xi\,, 
\eea
which gives
\be
\cL_a(\om^+) = -i a_c \,\Im z^c\, \om^+\,.\ee
Then 
\be
\cL_a(e^K) = 2a_c\Re z^c\, e^K\,,\quad\quad \cL_a\Bslb \Gbar_{B}dZ^B - \Zbar^A dG_A \Bsrb = -2 a_l x^l \bslb \Gbar_{B}dZ^B - \Zbar^A dG_A  \Bsrb 
 -i e^{-K_\Om}a_c dz^c 
\ee
which gives
\be
\cL_a(\om^3)=a_c \Im dz^c\,. 
\ee
The non-vanishing compensators from the duality symmetries are then
\be
W_{a}^\pm= 0\,,\quad\quad
W_{a}^3= \ha_c \Im z^c\,.
\ee

%%%%%%%%%%%%%%%%%%%%%%%%%%%%%%%%%%%%%%
\subsubsection{Hidden Symmetries}
%%%%%%%%%%%%%%%%%%%%%%%%%%%%%%%%%%%%%%
The hidden symmetries require more attention, they all have non-trivial compensators and the computation of these is somewhat intensive. As mentioned in the main text, a key to understanding the hidden symmetries is that the variation of the fields on the special K\"ahler base $\cM_z$ can be thought of as a $\xi$-dependent symmetry from section \ref{SymmKahler}. These parameters will now produce non-trivial terms when they appear under a derivative.

As an example we derive the variation of $\om^x$ under $k_{\eps_-}$. After some work we find the following expressions for $\eps_-$:
\bea
\delta_{\eps_-} (\om^+)&=&  i\blp - a_i y^i + e^{-2\phi} \brp \om^+ -\bslb i   2\sqrt{2} \, e^{\frac{K_\Om}{2}-\phi} (Z^T  \CC \xi) \bsrb  i \om^3-  d\Blp i2\sqrt{2} \, e^{\frac{K_\Om}{2}-\phi} Z^T \CC \xi \Brp 
\eea
Then comparing with \eq{LieSpinpm} we find that the compensator for our general Killing vectors is
\bea
W^{\pm}_{\epsilon_-}&=& - i 2\sqrt{2} \, e^{\frac{K_\Om}{2}-\phi} \, Z^T  \CC \xi \\
W^3_{\epsilon_-}&=&\ha_c \Im z^i- e^{-2\phi}
\eea
where
\be
\ha_c= -\half \blp 2 \xi^0 \txi_c-6 D_{cef} \xi^e \xi^f \brp
\ee
is the field-dependent parameter for the isometry on $\cM_z$.

%%%%%%%%%%%%%%%%%%%%%%%%%%%%%%%%%%%%%%
\section{Gaugings and their constraints} \label{app:gauging-constraints}
%%%%%%%%%%%%%%%%%%%%%%%%%%%%%%%%%%%%%%

For completeness the full set of constraints for the (symplectic) gaugings parameters are listed below.

The set of parameters
\begin{equation}
	\Theta^\cA = \{ \mathbb U, \alpha, \widehat\alpha^t, \epsilon_+, \epsilon_0, \epsilon_- \}
\end{equation} 
reads explicitly
\begin{equation}
	\mathbb U =
	\begin{pmatrix}
		\mathbb U^\Lambda \\
		\mathbb U_\Lambda
	\end{pmatrix}, \quad
	\alpha =
	\begin{pmatrix}
		\alpha^\Lambda \\
		\alpha_\Lambda
	\end{pmatrix},
	=
	\begin{pmatrix}
		\begin{pmatrix}
			\alpha^{A\Lambda} \\
			\alpha_A^\Lambda
		\end{pmatrix} \\
			\begin{pmatrix}
			\alpha^A_\Lambda \\
			\alpha_{A\Lambda}
		\end{pmatrix}
	\end{pmatrix}, \quad
	\widehat \alpha =
	\begin{pmatrix}
		\widehat \alpha^\Lambda \\
		\widehat \alpha_\Lambda
	\end{pmatrix},
	=
	\begin{pmatrix}
		\begin{pmatrix}
			\widehat \alpha^{A\Lambda} \\
			\widehat \alpha_A^\Lambda
		\end{pmatrix} \\
			\begin{pmatrix}
			\widehat \alpha^A_\Lambda \\
			\widehat \alpha_{A\Lambda}
		\end{pmatrix}
	\end{pmatrix}, \quad
	\begin{gathered}
		\epsilon_\pm =
		\begin{pmatrix}
			\epsilon_\pm^\Lambda \\
			\epsilon_{\pm\Lambda}
		\end{pmatrix}, \\
		\epsilon_0 =
		\begin{pmatrix}
			\epsilon_0^\Lambda \\
			\epsilon_{0\Lambda}
		\end{pmatrix}
	\end{gathered}
\end{equation} 
where $\mathbb U^\Lambda$ and $\mathbb U_\Lambda$ are matrices whose parameters depend on the model.

The constraints from the closure of the abelian algebra are
\begin{subequations}
\begin{itemize}
	\item electric/electric
		\begin{align}
			0 &= \mathbb T(\alpha_\Lambda, \hat\alpha_\Sigma) - \mathbb T(\alpha_\Sigma, \hat\alpha_\Lambda), \\
			0 &= - (\mathbb U_\Lambda \alpha_\Sigma
					- \mathbb U_\Sigma \alpha_\Lambda)
				+ (\epsilon_{0\Lambda} \alpha_\Sigma
					- \epsilon_{0\Sigma} \alpha_\Lambda)
				+ (\epsilon_{+\Lambda} \widehat\alpha_\Sigma
					- \epsilon_{+\Sigma} \widehat\alpha_\Lambda), \\
			0 &= (\mathbb U_\Lambda \widehat\alpha_\Sigma
					- \mathbb U_\Sigma \widehat\alpha_\Lambda)
				+ (\epsilon_{-\Lambda} \alpha_\Sigma
					- \epsilon_{-\Sigma} \alpha_\Lambda)
				+ (\epsilon_{0\Lambda} \widehat\alpha_\Sigma
					- \epsilon_{0\Sigma} \widehat\alpha_\Lambda), \\
			0 &= \alpha^t_\Lambda \mathbb C \alpha_\Sigma
				+ 2 (\epsilon_{+\Sigma} \epsilon_{0\Lambda}
					- \epsilon_{+\Lambda} \epsilon_{0\Sigma}), \\
			0 &= (\widehat\alpha^t_\Lambda \mathbb C \alpha_\Sigma
					- \alpha^t_\Lambda \mathbb C \widehat\alpha_\Sigma)
				+ 2 (\epsilon_{+\Sigma} \epsilon_{-\Lambda}
					- \epsilon_{+\Lambda} \epsilon_{-\Sigma}), \\
			0 &= \widehat\alpha^t_\Lambda \mathbb C \widehat\alpha_\Sigma
				+ 2 (\epsilon_{0\Lambda} \epsilon_{-\Sigma}
					- \epsilon_{0\Sigma} \epsilon_{-\Lambda}).
		\end{align}
	
	\item electric/magnetic
		\begin{align}
			0 &= \mathbb T(\alpha_\Lambda, \hat\alpha^\Sigma) - \mathbb T(\alpha^\Sigma, \hat\alpha_\Lambda), \\
			0 &= - (\mathbb U_\Lambda \alpha^\Sigma
					- \mathbb U^\Sigma \alpha_\Lambda)
				+ (\epsilon_{0\Lambda} \alpha^\Sigma
					- \epsilon_0^\Sigma \alpha_\Lambda)
				+ (\epsilon_{+\Lambda} \widehat\alpha^\Sigma
					- \epsilon_+^\Sigma \widehat\alpha_\Lambda), \\
			0 &= (\mathbb U_\Lambda \widehat\alpha^\Sigma
					- \mathbb U^\Sigma \widehat\alpha_\Lambda)
				+ (\epsilon_{-\Lambda} \alpha^\Sigma
					- \epsilon_-^\Sigma \alpha_\Lambda)
				+ (\epsilon_{0\Lambda} \widehat\alpha^\Sigma
					- \epsilon_0^\Sigma \widehat\alpha_\Lambda), \\
			0 &= \alpha^t_\Lambda \mathbb C \alpha^\Sigma
				+ 2 (\epsilon_+^\Sigma  \epsilon_{0\Lambda}
					- \epsilon_{+\Lambda} \epsilon_0^\Sigma), \\
			0 &= (\widehat\alpha^t_\Lambda \mathbb C \alpha^\Sigma
					- \alpha^t_\Lambda \mathbb C \widehat\alpha^\Sigma)
				+ 2 (\epsilon_+^\Sigma \epsilon_{-\Lambda}
					- \epsilon_{+\Lambda} \epsilon_-^\Sigma ), \\
			0 &= \widehat\alpha^t_\Lambda \mathbb C \widehat\alpha^\Sigma
				+ 2 (\epsilon_{0\Lambda} \epsilon_-^\Sigma 
					- \epsilon_0^\Sigma  \epsilon_{-\Lambda}).
		\end{align}
	
	\item magnetic/magnetic
		\begin{align}
			0 &= \mathbb T(\alpha^\Lambda, \hat\alpha^\Sigma) - \mathbb T(\alpha^\Sigma, \hat\alpha^\Lambda), \\
			0 &= - (\mathbb U^\Lambda \alpha^\Sigma
					- \mathbb U^\Sigma \alpha^\Lambda)
				+ (\epsilon_0^\Lambda \alpha^\Sigma
					- \epsilon_0^\Sigma \alpha^\Lambda)
				+ (\epsilon_+^\Lambda \widehat\alpha^\Sigma
					- \epsilon_+^\Sigma \widehat\alpha^\Lambda), \\
			0 &= (\mathbb U^\Lambda \widehat\alpha^\Sigma
					- \mathbb U^\Sigma \widehat\alpha^\Lambda)
				+ (\epsilon_-^\Lambda \alpha^\Sigma
					- \epsilon_-^\Sigma \alpha^\Lambda)
				+ (\epsilon_0^\Lambda \widehat\alpha^\Sigma
					- \epsilon_0^\Sigma \widehat\alpha^\Lambda), \\
			0 &= \alpha^{t\,\Lambda} \mathbb C \alpha^\Sigma
				+ 2 (\epsilon_+^\Sigma  \epsilon_0^\Lambda
					- \epsilon_+^\Lambda \epsilon_0^\Sigma), \\
			0 &= (\widehat\alpha^{t\,\Lambda} \mathbb C \alpha^\Sigma
					- \alpha^{t\,\Lambda} \mathbb C \widehat\alpha^\Sigma)
				+ 2 (\epsilon_+^\Sigma \epsilon_-^\Lambda
					- \epsilon_+^\Lambda \epsilon_-^\Sigma ), \\
			0 &= \widehat\alpha^{t\,\Lambda} \mathbb C \widehat\alpha^\Sigma
				+ 2 (\epsilon_0^\Lambda \epsilon_-^\Sigma 
					- \epsilon_0^\Sigma  \epsilon_-^\Lambda).
		\end{align}
\end{itemize}
\end{subequations}

We recall the expression of the matrix
\begin{equation} \label{TTensorDef}
	\mathbb T_{\alpha, \hat\alpha} = (\alpha^t \partial_\xi) (\hat\alpha^t \partial_\xi)\, \underline{S}.
\end{equation} 

The number of constraint from the algebra is
\begin{equation}
	\#(\text{algebra constraints}) = 3\, \frac{n_v (n_v - 1)}{2} \left[ \frac{n_h (n_h + 1)}{2} + 2 n_h + 3 \right]
\end{equation} 
where the $3$ comes from the three sets of constraints, the second front factor from the antisymmetric equations on $(\Lambda, \Sigma)$. The matrix $\mathbb T$ is symmetric.

The constraints from locality are
\begin{subequations}
\begin{align}
	0 &= \langle \alpha , \alpha^t \rangle = \alpha^\Lambda \alpha^t_\Lambda - \alpha_\Lambda \alpha^{t\Lambda}, \\
	0 &= \langle \alpha , \widehat\alpha^t \rangle = \alpha^\Lambda \widehat\alpha^t_\Lambda - \alpha_\Lambda \widehat\alpha^{t\Lambda}, \\
	0 &= \langle \widehat\alpha , \widehat\alpha^t \rangle =  \widehat\alpha^\Lambda \widehat\alpha^t_\Lambda - \widehat\alpha_\Lambda \widehat\alpha^{t\Lambda}, \\
	0 &= \langle \alpha , \epsilon_+ \rangle = \alpha^\Lambda \epsilon_{+\Lambda} - \alpha_\Lambda \epsilon_+^\Lambda, \\
	0 &= \langle \alpha , \epsilon_0 \rangle = \alpha^\Lambda \epsilon_{0\Lambda} - \alpha_\Lambda \epsilon_0^\Lambda, \\
	0 &= \langle \alpha , \epsilon_- \rangle = \alpha^\Lambda \epsilon_{-\Lambda} - \alpha_\Lambda \epsilon_-^\Lambda, \\
	0 &= \langle \widehat\alpha , \epsilon_+ \rangle = \widehat\alpha^\Lambda \epsilon_{+\Lambda} - \widehat\alpha_\Lambda \epsilon_+^\Lambda, \\
	0 &= \langle \widehat\alpha , \epsilon_0 \rangle = \widehat\alpha^\Lambda \epsilon_{0\Lambda} - \widehat\alpha_\Lambda \epsilon_0^\Lambda, \\
	\intertext{}
	0 &= \langle \widehat\alpha , \epsilon_- \rangle = \widehat\alpha^\Lambda \epsilon_{-\Lambda} - \widehat\alpha_\Lambda \epsilon_-^\Lambda, \\
	0 &= \langle \epsilon_+ , \epsilon_- \rangle = \epsilon_+^\Lambda \epsilon_{-\Lambda} - \epsilon_{+\Lambda} \epsilon_-^\Lambda, \\
	0 &= \langle \epsilon_+ , \epsilon_0 \rangle = \epsilon_+^\Lambda \epsilon_{0\Lambda} - \epsilon_{+\Lambda} \epsilon_0^\Lambda, \\
	0 &= \langle \epsilon_0 , \epsilon_- \rangle = \epsilon_0^\Lambda \epsilon_{-\Lambda} - \epsilon_{0\Lambda} \epsilon_-^\Lambda, \\
	0 &= \langle \mathbb U , \epsilon_+ \rangle = \alpha^\Lambda \epsilon_{+\Lambda} - \alpha_\Lambda \epsilon_+^\Lambda, \\
	0 &= \langle \mathbb U , \epsilon_0 \rangle = \alpha^\Lambda \epsilon_{0\Lambda} - \alpha_\Lambda \epsilon_0^\Lambda, \\
	0 &= \langle \mathbb U , \epsilon_- \rangle = \alpha^\Lambda \epsilon_{-\Lambda} - \alpha_\Lambda \epsilon_-^\Lambda, \\
	0 &= \langle \mathbb U , \alpha \rangle = \alpha^\Lambda \epsilon_{0\Lambda} - \alpha_\Lambda \epsilon_0^\Lambda, \\
	0 &= \langle \mathbb U , \widehat\alpha \rangle = \alpha^\Lambda \epsilon_{-\Lambda} - \alpha_\Lambda \epsilon_-^\Lambda
\end{align}
\end{subequations}
where
\begin{equation}
	\langle \alpha , \alpha^t \rangle =
	\begin{pmatrix}
		\langle \alpha^A , \alpha^B \rangle & \langle \alpha^A , \alpha_B \rangle \\
		\langle \alpha_A , \alpha^B \rangle & \langle \alpha_A , \alpha_B \rangle
	\end{pmatrix}, \qquad
	\langle \alpha , \epsilon_+ \rangle =
	\begin{pmatrix}
		\langle \alpha^A , \epsilon_+ \rangle \\
		\langle \alpha_A , \epsilon_+ \rangle
	\end{pmatrix}
\end{equation} 
and similarly for the others. The notation $\langle \mathbb U, X \rangle$ is shortcut for the product of $X$ with all parameters of $\mathbb U$ (by linearity). For example with a cubic prepotential one of the constraint is
\begin{equation}
	\langle \beta, X \rangle = 0, \qquad
	\beta =
	\begin{pmatrix}
		\beta^\Lambda \\ \beta_\Lambda
	\end{pmatrix}.
\end{equation} 

The numbers of locality constraints is
\begin{equation}
	\#(\text{locality constraints}) = 3 n_h^2 + 6 n_h + 3 + x\, n_h (2 n_h + 3).
\end{equation}

%%%%%%%%%%%%%%%%%%%%%%%%%%%%%%%%%%%%
\section{Black Hole Flow Equations}\label{sec:appendix-BH-eqs}
%%%%%%%%%%%%%%%%%%%%%%%%%%%%%%%%%%%%

In \cite{Halmagyi:2013sla} the equations for static black holes in electrically gauged $\cN=2$ supergravity were derived assuming that $P^1_\Lam=P^2_\Lam=0$. We can relax this assumption and still converge on the identical equations. Without any such assumption, with all components of $P^x_\Lam$ non-trivial in principle, the full set of equations for BPS black holes takes the form
\bea
p'^\Lam&=& 0 \label{pprime}\\
( p^\Lam P_\Lam^x)^2&=&\kappa^2  \label{qP1} \\
k_{\Lam}^u p^\Lam &=& 0 \\
\cL_r^\Lam P^x_{\Lambda} p^\Sig P_\Sig^x &=&  e^{2(V-U)} \Im ( e^{-i\psi}\cZ) \\
\del_r (e^U)&=&  \cL_i^\Lam P^x_{\Lambda} p^\Sig P_\Sig^x+ e^{2(U-V)}\Re ( e^{-i\psi}\cZ )   \\
\del_rV  &=&  2 e^{-U}  \cL^\Lam_i P^x_{\Lambda} p^\Sig P_\Sig^x  \\
\del_r  \blp e^{U} \cL_r^\Lam \brp  & =&\frac{1}{2 e^{2(V-U)}} \cI^{\Lam\Sig}\cR_{\Sig \Delta} p^\Delta -\frac{1}{2} \cI^{\Lam \Sigma } q_\Sigma  \label{delLr}\\
\del_r \blp e^{-U} \cL_i^\Lam \brp& =&\frac{p^{\Lam}}{2e^{2V}} + \frac{ 1}{2 e^{2U }}   \cI^{\Lam \Sig} P_\Sig^x p^\Delta P_\Delta^x  +\frac{4 }{e^{2U}}  \cL_r^\Sig P^x_\Sig p^\Delta P^x_\Delta  \cL_r^\Lam\\
  \dot{q}^u&=&2e^{-U} h^{uv}  \del_v \Blp p^\Sig P^x_\Sig \cL_i^{\Lam} P^x_\Lam\Brp  \label{qdot}
\eea
where we have defined the rescaled sections
\be
\cL^\Lam =\cL^\Lam_r+ i \cL_i^\Lam= e^{-i\psi} L^\Lam
\ee
and $e^{i\psi}$ is the phase of the supersymemtry parameter. First we define 
\be
P^x_p=P^x_\Lam p^\Lam
\ee
then use a local $SU(2)$ transformation to set 
\be
P_p^1=P^2_p=0\,,
\ee 
which is weaker than setting $P^1_\Lam=P^2_\Lam=0$ as was done in \cite{Halmagyi:2013sla}. At this point one can see that $P^1_\Lam$ and $P^2_\Lam$ completely drop out of the above equations \eq{pprime}-\eq{qdot}. This allows us to rewrite all equations in terms of $P^3_\Lambda \equiv P_\Lambda$ only. 

%%%%%%%%%%%%%%%%%%%%%%%%%%%%%%%%%%%%%%
\section{Killing Vectors on the Coset \texorpdfstring{$G/H$}{G/H}}\label{app:cosetconstruction}
%%%%%%%%%%%%%%%%%%%%%%%%%%%%%%%%%%%%%%
There is a general method to construct Killing vectors on cosets which we will utilize here \cite{Castellani:1983tb, Castellani:1999fz}. We first define a canonical automorphism on the Lie algebra of $G$:
\be
\tau(\vec{H})=-\vec{H}\,,\quad\quad \tau(E_i)=-F_i\,,\quad\quad \tau(F_i)=-E_i\,,\quad i=1,\ldots ,6\,.
\ee
We define a rotated basis $K_{\pm i}= E_i\pm F_i$
and the $\tau$-invariant subalgebra is given by $K_{-i}$
\be
\tau(K_{-i})=K_{-i}\,.
\ee

The general construction involves starting with a semi-simple Lie algebra $\ul{g}$ with generators $T_A$ and decomposing it into orthogonal subspaces under the Killing form 
\be
\kappa_{AB}= \Tr (T_AT_B)
\ee
as
\be
\ul{g}=\ul{h}+\ul{k}
\ee
The indices under this decomposition are $\{T_A\}=\{T_i,T_a\}$. The coset element is $L(y)$ and the one form has some component along $\ul{k}$ and some along $\ul{h}$:
\bea
V(y) &=& dL L^{-1} = V^a(y) T_a + \Om^i(y) T_i \\
V^a(y) &=& V^a_\al dy^\al
\eea
where we introduced coordinates $y^\al$ on the coset $G/H$.

We now produce a formula for the Killing vectors on $G/H$. If we vary the coset element by
\be
L\ra h L g
\ee
where
\bea
g&=& 1+ \eps^A T_A \\
h&=& 1- \eps^A W_A^i T_i \\
y'^\al&=& y^A+\eps^A K_A^\al(y)
\eea
we get
\be
\delta L(y)=  \eps^A K_A L(y)  =\eps^A \Bslb L T_A - W^i_A T_i L  \Bsrb\,.
\ee
From this we find that
\bea
L T_A L^{-1}&=&K_A^\al (\del_\al L)L^{-1} + W^i_A T_i 
\eea
which can be projected onto both $K$ and $H$ to give\footnote{We define the adjoint action to be \be g T_A g^{-1} = D_{A}^{\ B}T_B\ee}
\bea
D_A^{\ B}  \Tr \bslb T_B T_a\bsrb &=& K_A^\al\, \Tr \bslb (\del_\al L)L^{-1} T_a\bsrb \\
\Rightarrow\quad  \quad  D_A^{\ b} \kappa_{ba}&=&  K_A^\al V_{\al }^{\ b}\kappa_{ba}
\eea
and we now have an explicit formula for the Killing vectors on $G/H$ 
\be
K_A^\al=  D_A^{\ b} (V^{-1})_b^{\ \al}\,.
\ee
We now use this general formula to produce Killing vectors on the two canonical examples of homogeneous quaternionic K\"ahler manifolds.

%%%%%%%%%%%%%%%%%%%%%%%%%%%%%%%%%%%%%%
\section{\texorpdfstring{$G_{2(2)}/SO(4)$}{G2(2)/SO(4)}}\label{app:G2}
%%%%%%%%%%%%%%%%%%%%%%%%%%%%%%%%%%%%%%

We now construct the quaternionic K\"ahler metric on $G_{2(2)}/SO(4)$ and perform the explicit co-ordinate transformation such that the resulting space is clearly in the image of a c-map.

We take the following standard generators of $G_2$:
\begin{center}
$\begin{array}{ll}
H_1=\frac{1}{\sqrt{3}} \blp e_{11} - e_{22} + 2 e_{33} - 2 e_{55} + e_{66} - e_{77} \brp\,, & H_2 = e_{11}+e_{22}-e_{66} - e_{77}\,, \\
E_1=-2(e_{16}+ e_{27}) \,, & F_1=-\frac{1}{2} ( e_{61}+ e_{72}) \,, \\
E_2= \frac{1}{2\sqrt{3}} \blp 2e_{41}-2e_{52}-2e_{63}+2e_{74} \brp\,,& F_2= \frac{2}{\sqrt{3}} \blp e_{14}-e_{25}- e_{36}+e_{47} \brp\,, \\
E_3= \frac{1}{\sqrt{3}} \blp e_{13}-2e_{24} +2 e_{46} -2 e_{57} \brp\,, & F_3= \frac{1}{\sqrt{3}}\blp e_{31}-e_{42} + e_{64} - e_{75} \brp\,, \\
E_4=  -\frac{1}{\sqrt{3}} \blp e_{21} + e_{43} + e_{54} + e_{76}\brp\,,
 & F_4= -\frac{1}{\sqrt{3}} \blp e_{12} +2 e_{34}  +2 e_{45} + e_{67} \brp\,,  \\
E_5= \frac{1}{2} \blp e_{51}   + e_{73} \brp\,,
& F_5=-2(e_{15}- e_{37})\,, \\
E_6=-e_{23} -  e_{73} \,,
& F_6= - e_{32} - e_{65}\,.
\end{array}$
\end{center}
Using $K_{\pm i}= E_i \pm F_i$ we form the ordered basis
\be
T_A=\{ K_{-1}\,,\ldots , K_{-6},H_1,H_2,K_{+1},\ldots K_{+6} \}
\ee
the Cartan Killing Form is
\be
\kappa= 4 \bpm -1\!\!1_{6} & 0 \\ 0 & 1\!\! 1 _{8} \epm\,.
\ee

We take the coset element to be
\be
V=e^{ ( \vphi^1 H_1+\vphi_2 H_2 )/2}  e^{\zeta E_1} e^{\sqrt{3}(-\tha^1 E_2 +\tha^2 E_3)} e^{\ttha_2 E_6 }e^{2\sqrt{3}E_4} e^{-\ttha_1 E_5}
\ee
and the resulting metric is
\bea
ds^2&=& \frac{1}{4} \blp d\vphi_1^2+d\vphi_2^2 \brp + \sum_{i=1}^6 (\cF^i)^2 \,,
\eea
with the frames given by
\bea
\cF^1&=& d\zeta \,, \\
\cF^2&=& \sqrt{3} d\tha^1\,, \\
\cF^3&=& \sqrt{3} (d\tha^2-\zeta d\tha^1)\,, \\
\cF^4&=& 2\, 3^{1/2} \blp da+\half(\tha^1 d\tha^2 - \tha^2 d\tha^1) \brp\,, \\
\cF^5&=& d\ttha_1- 6 \tha^1 da-\tha^1(\tha^1 d\tha^2- \tha^2 d\tha^1)\,, \\
\cF^6&=& d\ttha_2 - 6 \tha^2 da-\tha^2(\tha^1 d\tha^2- \tha^2 d\tha^1) - \zeta \cF^5\,.
\eea
The co-ordinate transformation to bring the metric to the form \eq{quatmet} is
\bea
\vphi^1&=& -\sqrt{3} \Blp \vphi +\phi + \frac{\log(3)}{4}\Brp \\
\vphi^2&=&\phi -3 \vphi - \frac{3}{4} \log{3} \\
\zeta&=& 3^{3/4}\xi^0 \\
a&=& -\frac{1}{6 \, 3^{1/4}}\txi_1 -\frac{1}{2 \, 3^{1/4}} \chi \xi^1 \\
\tha^1&=&\frac{ \chi }{\sqrt{3}}\\
\tha^2 &=&  3^{1/4} (-\xi^1 +\chi \xi^0 ) \\
\ttha_1&=& \frac{1}{3^{3/4}}\Blp  \txi_0- \chi^2 \xi^1 \Brp \\
\ttha_2&=& -\sigma +2 \chi (\xi^1)^2 +\frac{1}{2} \xi^0 \txi^0 +\frac{1}{2} \xi_1 \txi_1 - \chi^2 \xi^0\xi^1 
\eea

The base special K\"ahler manifold is
\be
\cM_z= \frac{SU(1,1)}{U(1)} 
\ee
with prepotential
\be
\cG= -\frac{(Z^1)^3}{Z^0}
\ee
where the special co-ordinate on the $\cM_z$ is
\be
Z^A=\bpm1 \\ z \epm = \bpm 1 \\ \chi+  ie^{-2\vphi}\epm\,.
\ee
Explicitly the metric is the c-map coordinates is
\bea
ds^2&=& d\phi^2 + 3 d\vphi^2 + \frac{3}{4} e^{4\vphi} d\chi^2+ \frac{1}{4} e^{4\phi} \Blp d\sigma - \half \bslb \txi_A d\xi^A-\xi^A d\txi_A  \bsrb \Brp^2 \non \\
&& + \frac{1}{4} e^{2\phi-6\vphi} (d\xi^0)^2 + \frac{3}{4} e^{2\phi -2\vphi}(d\xi^1- \chi d\xi^0)^2 + \frac{1}{4}e^{2\phi+6\vphi}\blp d\txi_0 + \chi d\txi_1-\chi^3 d\xi^0 +3\chi^2 d\xi^1 \brp^2\non \\
&& + \frac{1}{12} e^{2\phi+2\vphi}\blp d\txi_1-3 \chi^2 d\xi^0 + 6\chi d\xi^1 \brp^2  \,.
\eea

We can also write down the Killing vectors in the quaternionic-K\"ahler construction in terms of those obtained from the coset construction $K_A$. First for the duality symmetries we find
\bea
h_{\eps_+}&=&\frac{1}{2} \bslb K_6+K_{14} \bsrb  \\
h_{\al_0}&=&\frac{3^{3/4}}{2} \bslb K_1+K_9  \bsrb \\
h_{\al_1}&=&-\frac{3^{3/4}}{2} \bslb K_3+K_{11}  \bsrb \\
h_{\al^0}&=&-\frac{1}{2\, 3^{3/4}}\bslb K_4+K_{12} \bsrb \\
h_{\al^1}&=&-\frac{1}{2\, 3^{3/4}}\bslb K_5+K_{13} \bsrb \\
h_{\eps_0} &=& \frac{\sqrt{3}}{2}\bslb K_7-\frac{1}{\sqrt{3}} K_8 \bsrb\\
h_a &=&  \frac{1}{2} \bslb K_2-K_{10} \bsrb  \\
h_{\beta}&=&-\frac{1}{2\, 3^{1/2}} \bslb K_7+\sqrt{3} K_8 \bsrb \\
h_b &=& -\frac{1}{2}\bslb K_2+K_{10} \bsrb
\eea
where $\{h_a,h_b,h_{\beta}\}$ refer to the obvious components of $h_{\UU}$. Then for the hidden symmetries we find
\bea
h_{\eps_-} &=& K_6-K_{14} \\
h_{\hal^0} &=& \frac{1}{3^{3/4}}\bslb K_4-K_{12} \bsrb\\
h_{\hal^1} &=&\frac{1}{3^{3/4}}\bslb K_5-K_{13} \bsrb \\
h_{\hal_0} &=&  \frac{1}{3^{3/4}}\bslb K_1-K_{9} \bsrb\\
h_{\hal_1} &=& \frac{1}{3^{3/4}} \bslb K_3-K_{11} \bsrb 
\eea

%%%%%%%%%%%%%%%%%%%%%%%%%%%%%%%%%%%%%%
\end{appendix}
%%%%%%%%%%%%%%%%%%%%%%%%%%%%%%%%%%%%%%
%%%%%%%%%%%%%%%%%%%%%%%%%%%%%%%%%%%%
\begin{appendix}
%%%%%%%%%%%%%%%%%%%%%%%%%%%%%%%%%%%%
\end{appendix}
%%%%%%%%%%%%%%%%%%%%%%%%%%%%%%%%%%%%%%%%%%%%%
%%%%%%%%%%%%%%%%%%%%%%%%%%%%%%%%%%%%%%%%%%%%%
\providecommand{\href}[2]{#2}\begingroup\raggedright\endgroup
\end{document}